\documentclass[twocolumn]{aastex63}

\newcounter{daggerfootnote}

\usepackage{xcolor}
\usepackage{longtable}
\usepackage{lipsum}
\usepackage{amsmath}
\usepackage[normalem]{ulem}
\usepackage[para]{threeparttable}
\usepackage{enumitem}
\usepackage[encapsulated]{CJK}
\usepackage{booktabs} 

\newcommand{\Msun}{\mbox{$M_\sun$}}
\newcommand{\Mstar}{$M_{\ast}$}
\newcommand{\muv}{M$_{\mathrm {UV}}$}

\newcommand{\hst}{\textit{HST}}

\newcommand\nion{$\dot n_{\rm ion}$}
\newcommand\fesc{$f_{\rm esc}$}
\newcommand\xiion{$\xi_{\rm ion}$}

\received{December 12, 2025}
\revised{February 20, 2026}
\accepted{February 28, 2026}
\published{April 20, 2026}

\submitjournal{\apj}

\begin{document}

\title{A GLIMPSE into the UV Continuum Slopes of the Faintest Galaxies in the Epoch of Reionization}

\correspondingauthor{Michelle C. Jecmen}
\email{mj32948@my.utexas.edu}
\author[0009-0004-4725-8559]{Michelle C. Jecmen}
\affiliation{Department of Astronomy, The University of Texas at Austin, Austin, TX 78712, USA}
\affiliation{Cosmic Frontier Center, The University of Texas at Austin, Austin, TX 78712, USA}

\author[0000-0002-0302-2577]{John Chisholm}
\affiliation{Department of Astronomy, The University of Texas at Austin, Austin, TX 78712, USA}
\affiliation{Cosmic Frontier Center, The University of Texas at Austin, Austin, TX 78712, USA} 

\author[0000-0002-7570-0824]{Hakim Atek}
\affiliation{Institut d'Astrophysique de Paris, CNRS, Sorbonne Universit\'e, 98bis Boulevard Arago, 75014, Paris, France}

\author[0000-0002-5588-9156]{Vasily Kokorev}
\affiliation{Department of Astronomy, The University of Texas at Austin, Austin, TX 78712, USA}
\affiliation{Cosmic Frontier Center, The University of Texas at Austin, Austin, TX 78712, USA} 

\author[0000-0003-4564-2771]{Ryan Endsley}
\affiliation{Department of Astronomy, The University of Texas at Austin, Austin, TX 78712, USA}
\affiliation{Cosmic Frontier Center, The University of Texas at Austin, Austin, TX 78712, USA} 

\author[0009-0009-9795-6167]{Iryna Chemerynska}
\affiliation{Institut d'Astrophysique de Paris, CNRS, Sorbonne Universit\'e, 98bis Boulevard Arago, 75014, Paris, France}

\author[0000-0001-6278-032X]{Lukas J. Furtak}
\affiliation{Department of Astronomy, The University of Texas at Austin, Austin, TX 78712, USA}
\affiliation{Cosmic Frontier Center, The University of Texas at Austin, Austin, TX 78712, USA} 

\author[0000-0002-9651-5716]{Richard Pan}
\affiliation{Department of Physics \& Astronomy, Tufts University, MA 02155, USA}

\author[0000-0001-7201-5066]{Seiji Fujimoto}
\affiliation{David A. Dunlap Department of Astronomy and Astrophysics, \\ University of Toronto, 50 St. George Street, Toronto, Ontario, M5S 3H4, Canada}
\affiliation{Dunlap Institute for Astronomy and Astrophysics, 50 St. George Street, Toronto, Ontario, M5S 3H4, Canada}

\author[0000-0003-3997-5705]{Rohan P.~Naidu}
\altaffiliation{NASA Hubble Fellow}
\affiliation{MIT Kavli Institute for Astrophysics and Space Research, 70 Vassar Street, Cambridge, MA 02139, USA}

\author[0000-0002-8984-0465]{Julian B.~Mu\~noz}
\affiliation{Department of Astronomy, The University of Texas at Austin, Austin, TX 78712, USA}
\affiliation{Cosmic Frontier Center, The University of Texas at Austin, Austin, TX 78712, USA}

\author[0000-0002-8192-8091]{Angela Adamo}
\affiliation{Department of Astronomy, The Oskar Klein Centre, Stockholm University, AlbaNova, SE-10691 Stockholm, Sweden}

\author[0000-0003-3983-5438]{Yoshihisa Asada}
\affiliation{David A. Dunlap Department of Astronomy and Astrophysics, University of Toronto, 50 St. George Street, Toronto, Ontario, M5S 3H4, Canada}
\affiliation{Dunlap Institute for Astronomy and Astrophysics, 50 St. George Street, Toronto, Ontario, M5S 3H4, Canada}

\author[0000-0001-8104-9751]{Arghyadeep Basu}
\affiliation{Univ Lyon, Univ Lyon1, Ens de Lyon, CNRS, CRAL UMR5574, F-69230, Saint-Genis-Laval, France}

\author[0000-0002-4153-053X]{Danielle A. Berg}
\affiliation{Department of Astronomy, The University of Texas at Austin, 2515 Speedway, Stop C1400, Austin, TX 78712, USA}
\affiliation{Cosmic Frontier Center, The University of Texas at Austin, Austin, TX 78712, USA} 

\author[0000-0003-1609-7911]{Jeremy Blaizot}
\affiliation{Univ Lyon, Univ Lyon1, Ens de Lyon, CNRS, CRAL UMR5574, F-69230, Saint-Genis-Laval, France}

\author[0000-0003-0348-2917]{Miroslava Dessauges-Zavadsky}  
\affiliation{Department of Astronomy, University of Geneva, Chemin Pegasi 51, 1290 Versoix, Switzerland}  

\author[0009-0004-3835-0089]{Emma Giovinazzo}
\affiliation{Observatoire de Genève, Université de Genève, Chemin Pegasi 51, 1290 Versoix, Switzerland}

\author[0000-0003-4512-8705]{Tiger Yu-Yang Hsiao}
\affiliation{Department of Astronomy, The University of Texas at Austin, Austin, TX 78712, USA}
\affiliation{Cosmic Frontier Center, The University of Texas at Austin, Austin, TX 78712, USA} 

\author[0000-0003-1561-3814]{Harley Katz}
\affiliation{Department of Astronomy \& Astrophysics, University of Chicago, 5640 S Ellis Avenue, Chicago, IL 60637, USA}
\affiliation{Kavli Institute for Cosmological Physics, University of Chicago, Chicago IL 60637, USA}

\author[0000-0002-3897-6856]{Damien Korber}
\affiliation{Observatoire de Genève, Université de Genève, Chemin Pegasi 51, 1290 Versoix, Switzerland}

\author[0000-0002-6149-8178]{Jed McKinney}
\altaffiliation{NASA Hubble Fellow}
\affiliation{Department of Astronomy, The University of Texas at Austin, Austin, TX 78712, USA}
\affiliation{Cosmic Frontier Center, The University of Texas at Austin, Austin, TX 78712, USA} 

\author[0000-0001-5538-2614]{Kristen.~B.~W. McQuinn}
\affiliation{Space Telescope Science Institute, 3700 San Martin Drive, Baltimore, MD, 21218, USA}
\affiliation{Rutgers University, Department of Physics and Astronomy, 136 Frelinghuysen Road, Piscataway, NJ 08854, USA} 

\author[0000-0001-5851-6649]{Pascal~A.~Oesch}  
\affiliation{Department of Astronomy, University of Geneva, Chemin Pegasi 51, 1290 Versoix, Switzerland} 
\affiliation{Cosmic Dawn Center (DAWN), Niels Bohr Institute, University of Copenhagen, Jagtvej 128, K\o benhavn N, DK-2200, Denmark}

\author[0000-0001-8419-3062]{A. Saldana-Lopez}
\affiliation{Department of Astronomy, Oskar Klein Centre, Stockholm University, 106 91 Stockholm, Sweden}

\author[0000-0001-7144-7182]{Daniel Schaerer}
\affiliation{Department of Astronomy, University of Geneva, Chemin Pegasi 51, 1290 Versoix, Switzerland}

%%%%%%%%%%%%%%%%%%%%%%%%%%%%%%%%%%%%%%%%%%%%%%%%%%%%%%%%%%%%%%%%%%%%%%%%%%%%%%%%%%%%%%%%%%%%%%%%%%%%%%%%%%%%%%%%%%%%%%%%%%%%%%%
\begin{abstract}
%%%%%%%%%%%%%%%%%%%%%%%%%%%%%%%%%%%%%%%%%%%%%%%%%%%%%%%%%%%%%%%%%%%%%%%%%%%%%%%%%%%%%%%%%%%%%%%%%%%%%%%%%%%%%%%%%%%%%%%%%%%%%%%

As observations have yet to constrain the ionizing properties of the faintest (\muv\ $\gtrsim -$16) galaxies, their contribution to cosmic reionization remains unclear. The rest-frame ultraviolet (UV) continuum slope ($\beta$) is a powerful diagnostic of stellar populations and one of the few feasible indicators of the escape fraction of ionizing photons (\fesc) for such faint galaxies at high-redshift. Leveraging ultra-deep JWST/NIRCam GLIMPSE imaging of strong lensing field Abell S1063, we estimate UV continuum slopes of 553 galaxies at z $>$ 6 with absolute magnitudes down to \muv\ $\simeq -$12.5. We find a modest evolution of $\beta$ with redshift and a flattening in the $\beta$-\muv\ relation such that galaxies fainter than \muv\ $\sim -$16.5 no longer exhibit the bluest UV slopes. The 136 ultra-faint galaxies with \muv\ $> -$16 are a diverse population encompassing dusty (30\%), old (15\%), and low-mass (50\%) galaxies. We apply the empirical $\beta$-\fesc\ relation from local Lyman continuum leakers, finding the mean \fesc\ peaks at $\sim 20\%$ at \muv$=-$16.5 and declines towards fainter galaxies, while remaining consistent with \fesc\ = 14\% within uncertainties, in agreement with recent radiative transfer simulations. Incorporating GLIMPSE constraints on the UV luminosity function, ionizing photon production efficiency, and escape fractions produces a reionization history consistent with independent observational constraints. Our results indicate galaxies with \muv\ between $-18$ and $-14$ supplied $\sim 60\%$ of the ionizing photons to cosmic reionization, while the lower \fesc\ of fainter galaxies produces a natural cutoff in the ionizing photon production rate density. 
\end{abstract}


%%%%%%%%%%%%%%%%%%%%%%%%%%%%%%%%%%%%%%%%%%%%%%%%%%%%%%%%%%%%%%%%%%%%%%%%%%%%%%%%%%%%%%%%%%%%%%%%%%%%%%%%%%%%%%%%%%%%%%%%%%%%%%%
\section{Introduction} \label{sec:intro}
%%%%%%%%%%%%%%%%%%%%%%%%%%%%%%%%%%%%%%%%%%%%%%%%%%%%%%%%%%%%%%%%%%%%%%%%%%%%%%%%%%%%%%%%%%%%%%%%%%%%%%%%%%%%%%%%%%%%%%%%%%%%%%%

Cosmic reionization marks the final major phase transition of the universe, during which hydrogen in the intergalactic medium (IGM) transforms from largely neutral to ionized. Reionization progresses as Lyman continuum (LyC) photons emitted by the first luminous sources ionize the surrounding hydrogen, creating expanding ionized bubbles that eventually coalesce to form a predominantly ionized universe by z $\sim$ 6 \citep{Fan2006,Planck2016,Robertson2022}. This process heats the IGM, regulates galaxy growth, and establishes the conditions for large-scale structure formation \citep{Gnedin2000, McQuinn2016}.

The timing and morphology of reionization depend sensitively on the abundance and spatial distribution of the sources which contributed the requisite ionizing photons \citep{Ouchii2009, Robertson2013, Madau2015, Robertson2015}. A reionization scenario dominated by numerous, evenly distributed sources would produce a relatively spatially uniform ionization field, whereas one driven by fewer, more clustered sources would result in a patchier, more inhomogeneous progression \citep{Finkelstein2019, Naidu2020}. This distinction is crucial, as reionization morphology determines subsequent galaxy formation and growth and impacts the cosmic microwave background (CMB) and Lyman-$\alpha$ forest measurements \citep[e.g.,][]{Mason2019}. 
 
The leading sources of reionization are thought to be star-forming galaxies \citep[e.g.,][]{ouchi2008, Robertson2013, Robertson2015, Bouwens2015, Finkelstein2019, Naidu2020} with active galactic nuclei (AGN) contributing at later times \citep[z $\simeq$ 5-7; e.g.,][]{Madau2015, Grazian2024, Madau2024, Singha2025}. However, each of these sources have their theoretical challenges. Star-forming galaxies are widespread and abundant but individually produce relatively few ionizing photons \citep[e.g.,][]{Begley2022}, many of which are absorbed by surrounding neutral gas and dust within the galaxy \citep{Gnedin2008,Bouwens2015UVLF}. In contrast, AGN are efficient ionizing sources \citep{Giallongo2015, Grazian2018} but are less common than star-forming galaxies and previously thought to only form well into reionization. 

Quantifying the contribution of star-forming galaxies to reionization requires constraining the ionizing emissivity (\nion), often expressed as
\begin{equation} 
    \dot n_{\rm ion} = \int \phi_{\rm UV}\ \xi_{\rm ion}\ L_{\rm UV}\ f_{\rm esc}\ dM_{\rm UV}
\end{equation}
where $\phi_{\rm UV}$ is the UV luminosity function (UVLF), $\xi_{\rm ion}$ is the intrinsic production rate of ionizing photons per UV continuum luminosity ($L_{\rm UV}$), and \fesc\ is the fraction of LyC photons which escape into the IGM. Each of these factors is expected to vary with both redshift and galaxy properties (e.g. absolute magnitude, stellar mass, star-formation rate).

Faint, low-mass galaxies have long been compelling candidates for dominating \nion\ during the Epoch of Reionization (EoR). In hierarchical structure formation, such galaxies are abundant at early times, driving a steep increase in $\phi_{\rm UV}$ toward faint magnitudes \citep{Bouwens2014, Finkelstein2015, Atek2015}. Their low metallicities and bursty star-formation histories yield high $\xi_{\rm ion}$ \citep[e.g.,][]{Endsley2023, Prieto2023}, while their low dust content and shallow gravitational potentials enable stellar feedback to efficiently remove neutral hydrogen to promote higher \fesc\ than in massive systems \citep{Trebitsch2017, Rosdahl2022}. Taken together, these properties imply that large populations of faint galaxies at high redshift intrinsically produce substantial ionizing radiation and efficiently leak it into the IGM, thereby powering reionization. These faint galaxies have long been theorized as the dominant contributors to reionization, but have always been beyond observational constraints \citep{Finkelstein2012b, Robertson2013}.

Initial JWST observations suggest galaxies in the early universe were more prolific producers of ionizing photons than previously anticipated. The overabundance of bright galaxies at early times results in minimal evolution in the bright-end of the galaxy UV luminosity function \citep[e.g.][]{Naidu2022,Finkelstein2023, Harikane2023, Donnan23}. In addition, initial measurements of $\xi_{\rm ion}$ at z $>$ 5 revealed a population of extremely efficient producers of ionizing photons \citep{Endsley2023, Prieto2023, Simmonds2023, Atek2024}, however subsequent analysis has revealed a more complex picture where low-mass galaxies undergo extremely episodic ``bursty" star formation histories \citep{Pahl2023, Endsley2024, Simmonds2024}. Collectively these findings, when extrapolated to include fainter galaxies, imply a much larger ionizing emissivity which produces a reionization history that evolves too rapidly compared to constraints from the CMB and Lyman-$\alpha$ forest \citep{Munoz2024}.

To reconcile this tension, several explanations have been proposed: over-predicted average $\xi_{\rm ion}$ values from bursty low-mass galaxies \citep{Simmonds2024, Pahl2023}, overestimated \fesc\ in faint galaxies \citep{Papovich2025}, or a higher IGM clumping factor \citep{Davies2024}. However, the faintest galaxies (hereafter defined as \muv $\gtrsim -16$) at z $>$ 6 remain largely unobserved, leaving their true contribution to reionization uncertain. 

A key challenge is constraining \fesc. Direct measurements of \fesc\ at z $\gtrsim$ 4 are statistically unlikely due to absorption from neutral hydrogen in the IGM \citep{Inoue2014}. Instead, local analogs of LyC leakers have been studied in-depth, suggesting that galaxies with high \fesc\ exhibit minimal line-of-sight interstellar medium (ISM) absorption, high ionization emission-lines, and compact star-formation \citep{Izotov2016a, Izotov2016b, Izotov2018a, Izotov2018b, Gazagnes2018, Steidel18, SaldanaLopez2022, Flury2022, Flury2022b, Pahl2023, Jaskot2024a, Jaskot2024}. Several diagnostics have been proposed based on these trends, including small Ly$\alpha$ peak separations \citep{Verhamme2015, Gazagnes2020}, compact Ly$\alpha$ halos \citep{SaldanaLopez2025}, \ion{Mg}{2} emission \citep{Henry2018, Chisholm2020, Xu2022, Gazagnes2025}, high [\ion{O}{3}]/[\ion{O}{2}] ratio \citep{Izotov2018a}, and blue UV continuum slopes \citep{Chisholm2022}.

The UV continuum slope ($\beta$, where $F_{\lambda} \propto \lambda^{\beta}$) is a promising population-level probe of \fesc\ in faint, high-redshift galaxies. A galaxy's intrinsic UV slope is set by the massive star population and limited to blue values between $-3.0$ and $-2.5$ for stellar populations younger than 50 Myr, varying according to metallicity and age \citep[e.g.][]{Schaerer2005Stellar-populat,Katz2025a}. The observed UV slope is reddened by the addition of nebular continuum to $\beta \gtrsim -2.6$, and can reach $\beta \sim -1.0$ in extreme nebular continuum systems \citep{Katz2025a}. Typically, the UV slope traces and is reddened most substantially by dust attenuation, producing values of $\beta$ $\sim -2.0$ with even small E(B-V) $\sim$ 0.04 mag. As such, red UV continuum slopes indicate the presence of dust, which is an efficient tracer of cool gas and strongly absorbs ionizing photons. Thus, redder UV slopes have been empirically shown at low-redshift to correlate with lower \fesc\ values \citep{Chisholm2022}. In contrast, UV slopes bluer than $\beta \sim -2.6$ cannot be reddened by dust and gas, suggesting high \fesc\ \citep[e.g.][]{Topping2022,Cullen2024}. 

The UV continuum slope has been widely studied for galaxies between z $\simeq 2-9$ with both HST \citep[e.g.][]{Dunlop2012, Finkelstein2012, Rogers2013, Dunlop2013, Bouwens2014, Morales2023, Bhatawekar2021} and JWST \citep[e.g.][]{Cullen2023, Nanayakkara2023, Morales2024, Topping2024, Austin2024, Saxena2024, Donnan2025}. In general, the population average $\beta$ becomes bluer at higher redshifts, decreasing from $\langle \beta \rangle =-1.85$ at z$\sim$4 to $\langle \beta \rangle =-2.05$ at z$\sim$7 for the $\langle \rm{M_{UV}} \rangle \sim -19.5$ HST-compiled sample of \citet{Bouwens2014}, likely reflecting reduced dust content or grayer attenuation curves \citep{McKinney2025}. Many studies find $\beta$ also decreases towards fainter galaxies and/or lower stellar masses \citep{Bouwens2012, Bhatawekar2021, Nanayakkara2023, Saldana-Lopez2023,Topping2024, Dottorini2025}, although these measurements are biased by photometric scatter and have yet to probe galaxies fainter than \muv $\sim -16$ \citep{Cullen2023,Austin2024,Morales2024}. 

This work aims to use UV continuum slopes to provide the first empirical population-level estimates of the escape fractions of galaxies fainter than \muv $\sim -16$ in the EoR. While the most robust \fesc\ estimates require a multivariate approach, \citet{Jaskot2024} shows that the UV slope and neutral gas tracers share the most predictive power for \fesc. The UV slope allows us to test the long-standing assumption that the faintest galaxies exhibit the highest escape fractions as a starting point to understanding their contribution to cosmic reionization.

Throughout this work we analyze data from the JWST GLIMPSE survey \citep{Atek2025}, a program which leverages ultra-deep imaging and strong gravitational lensing to estimate UV slopes from the faintest galaxies in the EoR. In Section \ref{sec:obs_red}, we describe the observations and sample selections. Section \ref{sec:uv_slopes} outlines our methods for measuring $\beta$ and we present our results for how $\beta$ evolves with redshift and absolute magnitude in Section \ref{sec:res}. We present the diversity of faint galaxies in our sample and evidence for extremely blue UV slopes, followed by a discussion of the estimated escape fractions and their implications on cosmic reionization in Section \ref{sec:disc}. We conclude with a summary of key findings in Section \ref{sec:conclude}. Throughout this paper we assume a flat $\Lambda$CDM cosmology with $H_0 =70\ \mathrm{km\ s^{-1}}$, $\Omega_{\mathrm m} = 0.3$, and $\Omega_{\mathrm \Lambda} = 0.7$. All magnitudes are AB. 

%%%%%%%%%%%%%%%%%%%%%%%%%%%%%%%%%%%%%%%%%%%%%%%%%%%%%%%%%%%%%%%%%%%%%%%%%%%%%%%%%%%%%%%%%%%%%%%%%%%%%%%%%%%%%%%%%%%%%%%%%%%%%%
\section{Observations and Reductions} \label{sec:obs_red}
%%%%%%%%%%%%%%%%%%%%%%%%%%%%%%%%%%%%%%%%%%%%%%%%%%%%%%%%%%%%%%%%%%%%%%%%%%%%%%%%%%%%%%%%%%%%%%%%%%%%%%%%%%%%%%%%%%%%%%%%%%%%%%

Here we briefly describe the observations, data reductions, and sample selection. A fully detailed breakdown of these steps is given in \citet{Atek2025}. 

This work uses imaging data from the Cycle 2 JWST GLIMPSE survey (GO-3293, PIs: H. Atek \& J. Chisholm). The program leverages ultra-deep imaging and strong gravitational lensing of the massive foreground cluster Abell S1063 to probe the faintest, high-redshift galaxies with $\sim$155 total hours across 7 broadband (F090W, F115W, F150W, F200W, F277W, F356W, F444W) and 2 medium-band (F410M, F480M) filters. All broadband images reach 5$\sigma$ depths in 0\farcs2 apertures of $\sim$30.8 mag, requiring up to 40 hours of integration time. Combined with gravitational lensing, we reach intrinsically faint magnitudes down to \muv\ $\sim -12.5$ with magnification $\mu \sim 34$ at $z \sim 6.7$. 

The data reduction follows the procedure of \citet{Endsley2024} using the \textit{JWST} Science Calibration Pipeline and Calibration Reference Data System (CRDS) files from \textsc{jwst\_1293.pmap}. This procedure includes additional corrections for cosmic rays, 1/f noise, wisp removal, artifact correction, and stray light subtraction. We construct our own flat field images using all public NIRCam images as of January 12, 2025 to remove correlated noise as a result of sub-pixel dithering, recovering up to 0.5 mag in depth for the long-wavelength filters. Following the methods of \citet{Shipley2018} and \citet{Weaver2024}, bright cluster galaxies (bCGs) and intracluster light (ICL) are modeled and subtracted from the image.

All images are convolved to the Point Spread Functions (PSF) of the F480M data using PSFs empirically derived using stars within the GLIMPSE NIRCam field. Sources are detected in inverse-variance weighted stacks of both short-wavelength (SW) channels (F090W, F115W, F150W, and F200W) and  long-wavelength (LW) channels (F277W, F356W, and F444W). Both detection images crucially explore the UV properties of distant galaxies as extremely blue high-redshift galaxies might only be detected in the SW stacks.  We use \textsc{source extractor} \citep{Bertin1996} to define the locations of sources and then merge the SW and LW catalogs into a single catalog. We measure flux densities with 0.2" circular apertures, appropriate for the faint high-redshift sources.

Sources are selected with two complementary techniques. The first focuses on a narrow redshift range between $z \sim 6.1-6.6$ where the H$\alpha$ line is redshifted into the F480M filter, the strong [\ion{O}{3}] lines are in the F356W filter, and the F410M samples continuum regions. At $z=6.1-6.6$, we do not have a JWST Lyman Break filter (e.g. F070W) and the available \textit{HST} data do not cover the entire GLIMPSE field. This means that SED fitting relies on strong rest-frame optical features. We first fit the entire GLIMPSE catalog with \textsc{eazy} using the standard \texttt{blue\_sfhz\_13} models while also including tailored \textsc{beagle} models to sample older stellar populations (see Chisholm et al. in prep.). For every $z = 6.1-6.6$ galaxy that \textsc{eazy} selects we also fit with \textsc{beagle} and require both codes to have a $>95$\% probability that the sources are between redshifts of 5-7. We then visually inspect all photometric candidates to remove strong artifacts, sources near edges of the detectors, or individual strongly lensed clumps within galaxies. After all of these cuts, we are left with 94 photometrically selected galaxies at $z=6.1-6.6$, but only 71 of these sources are \textit{not} also selected by the Lyman break selection (see below). In Chisholm et al.\ (in prep.) we compute the completeness of the photo$-z$ selection, finding that we are highly complete down to \muv $\sim -15$, and this completeness does not significantly depend on the UV continuum slope. We refer the reader to Chisholm et al.\ (in prep.) for more details on this sample selection. 

The second selection technique uses a combination of the Lyman break technique and \textsc{eazy} photometric fitting \citep[e.g., ][]{Atek2023}. We use a color-color selection criterion that is separately defined for four redshift bins (6 $<$ z $<$ 9; 9 $<$ z $<$ 11; 11 $<$ z $<$ 15; z$>$15) and requires a Lyman Break of at least 0.8~mag. We then fit the photometry of all Lyman Break selected sources with \textsc{easy} to confirm the redshift, and visually inspect each source. These sources are used to calculate the GLIMPSE UVLF at z $>$ 9 \citep{Chemerynska2025} and z $<$ 6 (Atek et al.\ in prep.), and their associated completeness estimates apply to this subsample. We refer the reader to \citet{Kokorev2025}, \citet{Chemerynska2025}, and \citet{Atek2015} for additional selection criteria and detection thresholds. In total, we are left with a full sample of 553 sources across redshifts from 6-16.

We use a new strong lensing model of the lensing cluster AS1063, following the methodology outlined by \citet{Furtak2023} and using the updated version of the \citet{Zitrin2015} parametric code, sometimes referred to as \texttt{Zitrin-analytic}. This is a fully analytic method, such that the model is not limited to a grid resolution. The total mass distribution of AS1063 is modeled as two smooth dark matter (DM) halos \citep[following e.g.][]{Bergamini2019,Beauchesne2024}, which are parametrized as pseudo-isothermal elliptical mass distributions \citep[PIEMDs;][]{Kassiola1993}, and 303 cluster galaxies which are parametrized as dual pseudo-isothermal ellipsoids \citep[dPIEs;][]{Eliasdottir2007}. We use 75 multiple images of 28 sources to constrain the lens model, which achieves an average lens plane image reproduction error of $\Delta_{\mathrm{RMS}}=0.54\arcsec$. For more details on the lens model, its methods, constraints, and results, we refer the reader to Furtak et al.\ (in prep.). The final lens model produces magnification values at each object's coordinates and redshift, which we use to correct the observed magnitudes, stellar masses, and star-formation rates.

%%%%%%%%%%%%%%%%%%%%%%%%%%%%%%%%%%%%%%%%%%%%%%%%%%%%%%%%%%%%%%%%%%%%%%%%%%%%%%%%%%%%%%%%%%%%%%%%%%%%%%%%%%%%%%%%%%%%%%%%%%%%%%%
\section{UV Slopes} \label{sec:uv_slopes}
%%%%%%%%%%%%%%%%%%%%%%%%%%%%%%%%%%%%%%%%%%%%%%%%%%%%%%%%%%%%%%%%%%%%%%%%%%%%%%%%%%%%%%%%%%%%%%%%%%%%%%%%%%%%%%%%%%%%%%%%%%%%%%%

%--Fig1: Grid of SEDs showing UV slope calculation & range
\begin{figure*}[]
    \includegraphics[width=\textwidth]{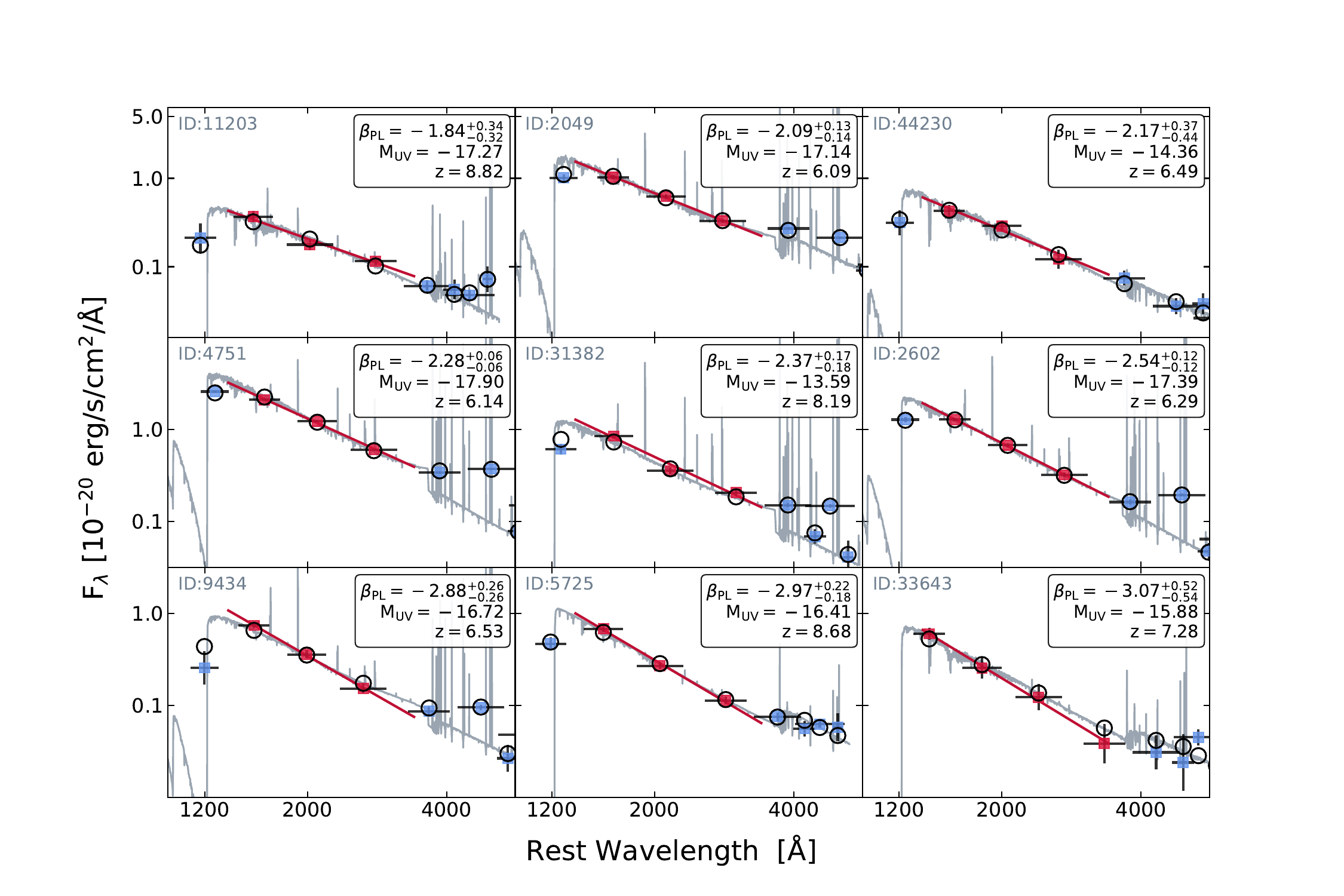}
    \caption{Representative galaxies ranging in UV slope ($\beta_{\rm PL}$), absolute magnitude (\muv), and photometric redshift (z) to illustrate the calculation of $\beta_{\rm PL}$, ordered from reddest to bluest. Square points are observed photometry, colored red if used in power-law fitting (red line) and blue otherwise. The model SEDs are each shown in their rest-frame as gray curves with the predicted photometry as empty black circles. Most models can reproduce both the UV continuum and optical emission features, except for sources with extremely blue $\beta_{\rm PL}$ (bottom row).}
    \label{fig:betagrid}
\end{figure*}

For all 553 sources, we fit \textsc{Bagpipes} \citep{Carnall2018} spectral energy distribution (SED) models to all seven broadband and two medium band filters. We adopt a constant star formation history, allowing galaxy ages to vary between 10-100 Myr. In addition, we include a young burst component with an age range of 1–10 Myr to approximate a simple non-parametric star formation history, allowing for bursty star formation theorized at high redshift. For each star formation history, the observed stellar mass formed is limited to $\log(M_{\star}) \in [4, 11]$, which we then de-magnify to account for gravitational lensing. We apply dust attenuation following the Calzetti dust law \citep{Calzetti2000} with a log prior extinction coefficient (A$_{\rm V}$) in the range [1e-5, 2]. We adopt uniform priors on metallicity and ionization parameter within the ranges $Z/Z_\odot \in [0.001, 2.5]$ and $\log(U) \in [-3, -1]$, respectively. Reported photometric redshifts are adopted from the maximum-likelihood \textsc{Bagpipes} model and uncertainties from the inner 68th percentile of the posterior distribution. Figure \ref{fig:betagrid} shows the maximum-likelihood SEDs for representative galaxies which span the range of UV continuum slopes in our sample.

We measure the UV slope ($\beta_{\rm PL}$) by fitting a power law ($f_{\lambda} \propto \lambda^{\beta}$; \citealt{Calzetti1994}) to the rest-frame UV photometry, similar to \citet{Dunlop2013, Rogers2014, Topping2024}. We highlight the filters used and the resulting best-fit line from this fitting procedure in red for each source in Figure \ref{fig:betagrid}. We fit $\beta_{\rm PL}$ using all photometric filters with rest-frame pivot wavelengths in the range 1350Å - 3400Å, a window chosen to avoid Ly$\alpha$ and contamination from rest-frame optical nebular emission lines and stellar continuum features while maximizing filter coverage across redshift. The filters that cover the defined window shift systematically with redshift, as reported in Table \ref{tab:filters}. Roughly 90\% of our sources are fit using three filters.

%--Tab1: Filters used in UV slope calculation for each galaxy redshift
\begin{table}[h]
    \centering
    \caption{Filters used in calculating the UV slope for each galaxy redshift.}
    \label{tab:filters}
    \setlength{\tabcolsep}{3pt}
    \begin{tabular}{lc}
        \hline
        Redshift & Filters used to fit $\beta_{\rm PL}$ \\
        \hline
        $5.7 \leq z < 7.1$   & F115W, F150W, F200W \\
        $7.1 \leq z < 7.6$  & F115W, F150W, F200W, F277W \\
        $7.6 \leq z < 9.5$ &  F150W, F200W, F277W \\
        $9.5 \leq z < 10.1$ &  F150W, F200W, F277W, F356W \\
        $10.1 \leq z < 11.1$ &  F200W, F277W, F356W \\
        $11.1 \leq z < 12$ &  F200W, F277W, F356W, F410M \\
        $12 \leq z < 13.2$ &  F200W, F277W, F356W, F410M, F444W  \\
        %$13.2 \leq z < 13.8$ &  F200W, F277W, F356W, F410M, F444W, F480M  \\ no galaxies in this bin
        $13.2 \leq z$ &  F277W, F356W, F410M, F444W, F480M  \\
        \hline
    \end{tabular}
\end{table}

Uncertainties on $\beta_{\rm PL}$ are estimated via Monte Carlo resampling. For each galaxy, we perturb both the redshift and photometry within errors and refit $\beta$ over 1000 realizations. The 16th and 84th percentiles of the resulting $\beta$ distribution define the lower and upper uncertainties, respectively. As discussed in \citet{Austin2024}, calculating the UV slope directly from broadband photometry can be biased by the inclusion of UV line emission and photometric ``up-scattering" of filters red-ward of the Lyman break into the detection threshold (see Section \ref{sec:beta-muv} for a discussion of how this bias affects our sample). As such, we validate our $\beta$ measurement by additionally estimating $\beta_{\rm SED}$ by fitting a power-law to the rest-frame UV continuum of the maximum-likelihood \textsc{Bagpipes} SED. As advocated by \citet{Finkelstein2012}, this method utilizes all nine filters and thus encodes optical nebular and stellar emission, as shown by the gray best-fit SED curves in Figure \ref{fig:betagrid}. These features constrain the massive star population capable of producing ionizing photons and can mitigate photometric scatter biasing $\beta_{\rm PL}$ toward unphysical blue values. For this measurement, we adopt the narrower rest-frame wavelength window of 1350Å-1800Å to more closely align with \citet{Chisholm2022} and apply a 2$\sigma$ clipping of spectral pixels in the best-fit SED model to remove UV emission lines. We calculate $\beta_{\rm SED}$ across all 500 posterior samples from \textsc{Bagpipes}, reporting the 16th and 84th percentiles of distribution as the uncertainties.

\begin{figure}
    \centering
    \includegraphics[width=\linewidth]{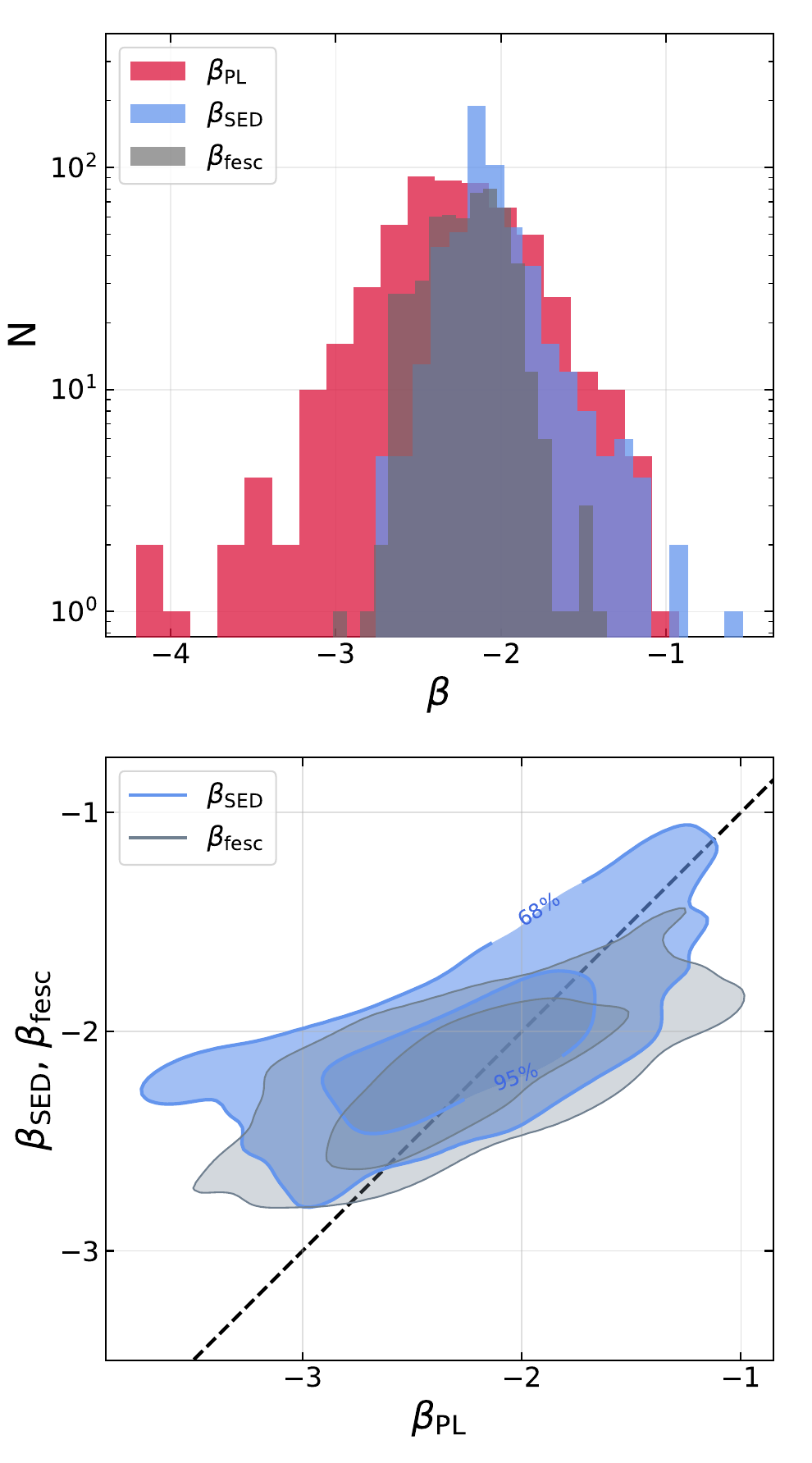}
    \caption{Comparison of UV slope measurements from power-law fits to photometry ($\beta_{\rm PL}$) and to model SEDs with \fesc=0 ($\beta_{\rm SED}$) and with varying \fesc\ ($\beta_{\rm fesc}$). 
    \textbf{Upper panel:} 
    Histograms for each $\beta_{\rm PL}$ (red), $\beta_{\rm SED}$ (blue), and $\beta_{\rm fesc}$ (gray), with $\beta_{\rm PL}$ showing the widest distribution.     
    \textbf{Lower panel:} 
    Comparing the agreement of $\beta_{\rm PL}$ to the UV slope inferred from model SEDs for both $\beta_{\rm SED}$ (blue contours) and $\beta_{\rm fesc}$ (gray contours). Outer and inner contours enclose the 68th and 95th percentiles. The dashed black line indicates the one-to-one relation, from which contours deviate at very blue UV slopes not reproduced by models.}
    \label{fig:betabeta}
\end{figure}

Figure \ref{fig:betabeta} compares the distribution of UV slopes as measured from power-law fitting to photometry ($\beta_{\rm PL}$, red) versus from the model SEDs ($\beta_{\rm SED}$, blue). Most notably, $\beta_{\rm PL}$ has a much wider distribution which peaks at slightly bluer UV slopes ($\langle \beta_{\rm PL} \rangle \sim -2.27$ compared to $\langle \beta_{\rm SED} \rangle \sim -2.06$). As expected, we find both methods yield consistent values within 1$\sigma$ for 64\% of sources in the sample, primarily for $\beta > -2.5$ as illustrated by the alignment of the blue contour curves with the one-to-one ($\beta_{\rm PL}$ = $\beta_{\rm SED}$) line in Figure \ref{fig:betabeta}. We find that SED fitting does not reproduce the very blue values of $\beta_{\rm PL} < -2.75$ measured directly from photometry for even the youngest, metal-poor, and dust-free stellar populations (bottom row of Figure \ref{fig:betagrid}). This observed lower limit on $\beta_{\rm SED}$ is responsible for the deviation of the blue contours away from the one-to-one line. 

We test if this lower limit on $\beta_{\rm SED}$ arises from the \textsc{Bagpipes} requirement of zero ionizing photon escape, as explored in \citet{Morales2024, Austin2024, Giovinazzo2025}. Even in dust-free models, ionizing photons must produce nebular continuum emission that reddens the UV slope, imposing a $\beta_{\rm SED}$ floor. To explore whether allowing for a non-zero escape fraction can reproduce our observed very blue $\beta_{\rm PL}$, we additionally fit the adapted \textsc{Bagpipes} model of \citet{Giovinazzo2025} which implements a simple ``picket-fence" model of LyC escape. From these SEDs, we calculate $\beta_{\rm fesc}$ using the same procedure as $\beta_{\rm SED}$, adopting a uniform prior on \fesc\ as to reproduce $\beta_{\rm fesc} < -2.75$. 

Adding \fesc\ as a free parameter results in a systematically bluer ($\langle \beta_{\rm fesc} \rangle \sim -2.20$) distribution compared to $\beta_{\rm SED}$, as shown by the gray histogram and contours in Figure \ref{fig:betabeta}. We find that $\beta_{\rm fesc}$ and $\beta_{\rm PL}$ agree within 1$\sigma$ for 78\% of sources and that the adapted model can reproduce a UV continuum as blue as $\beta_{\rm fesc} = -3.0$. However, we see that $\beta_{\rm fesc}$ is bluer than $\beta_{\rm PL}$ for galaxies with relatively red UV slopes, causing a worse agreement of the gray contour with the one-to-one line for red slopes in Figure \ref{fig:betabeta}. We further explore the impact of non-zero LyC escape on UV slopes in Section \ref{sec:extblue}, where we examine our subset of galaxies with extremely blue UV slopes.

The UV slopes estimated from both SED models typically agree with our $\beta_{\rm PL}$ measurements within 1$\sigma$ uncertainties, although neither method reproduces the full distribution inferred from photometry. We adopt $\beta_{\rm PL}$ as our fiducial UV slope estimation, as this approach does not restrict the range of allowable values while allowing uncertainties to reflect photometric noise. Individual measurements may be scattered to unphysically blue values which cannot be explained by current SED modeling; however, this scatter is expected for intrinsically blue populations and is essential to fully sample in order to recover unbiased average UV slopes. Throughout, we use the SED fits to test our conclusions, but overall they are largely insensitive to this choice of methodology. A subset of our sample and its derived properties are given in Table \ref{tab:catalogue} and the full table is available online.

%--Fig2: UV slope from PL vs. SED distribution

%%%%%%%%%%%%%%%%%%%%%%%%%%%%%%%%%%%%%%%%%%%%%%%%%%%%%%%%%%%%%%%%%%%%%%%%%%%%%%%%%%%%%%%%%%%%%%%%%%%%%%%%%%%%%%%%%%%%%%%%%%%%%%%
\section{Results} \label{sec:res}
%%%%%%%%%%%%%%%%%%%%%%%%%%%%%%%%%%%%%%%%%%%%%%%%%%%%%%%%%%%%%%%%%%%%%%%%%%%%%%%%%%%%%%%%%%%%%%%%%%%%%%%%%%%%%%%%%%%%%%%%%%%%%%%

\subsection{Redshift Evolution of UV Slopes}\label{sec:beta-z}

\begin{figure}[h]
    \centering
    \includegraphics[width=\linewidth]{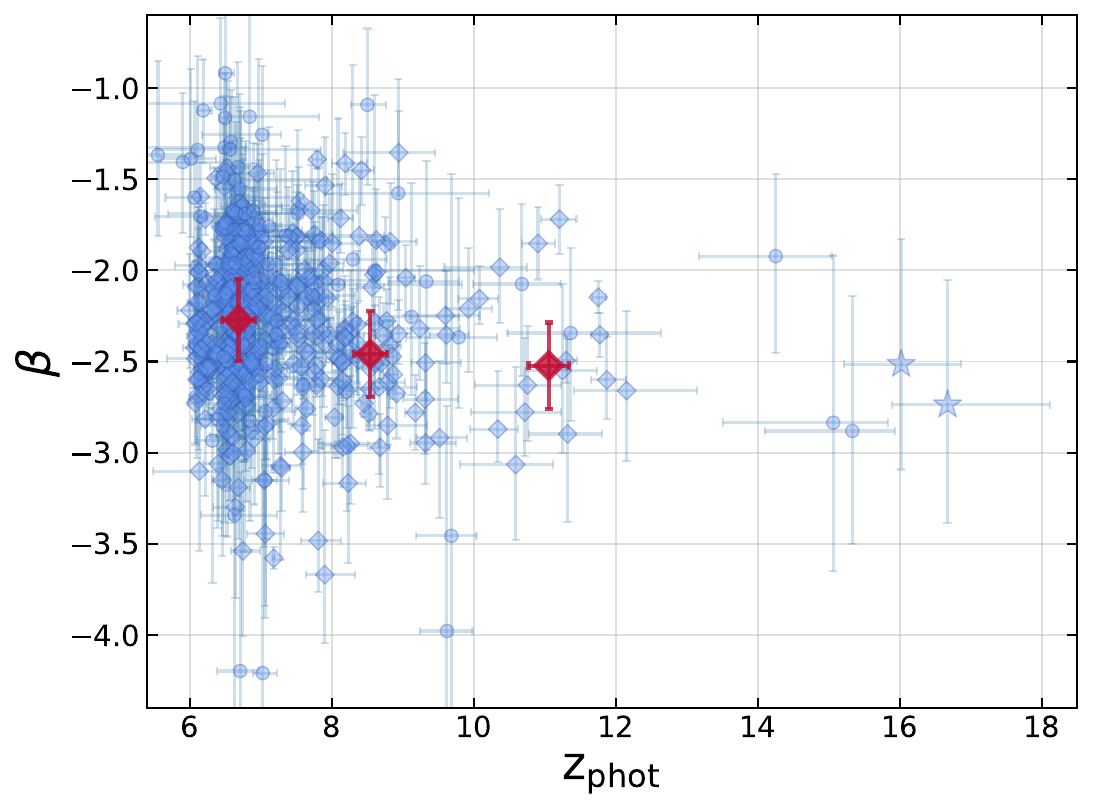}
    \caption{Mild evolution of $\beta$ with redshift driven by a lack of red galaxies at high redshift. Red diamonds are median bins for galaxies with robust (S/N $\geq$ 5) $\beta$ measurements, plotted as blue diamonds. The two highest redshift sources (blue stars) are the high-z galaxy candidates of \citet{Kokorev2025}.}
    \label{fig:beta-z}
\end{figure}

In Figure \ref{fig:beta-z} we show the evolution of $\beta$ with photometric redshift for our full sample. To understand how the UV slope of a typical galaxy changes with redshift, we calculate the median redshift and UV slope of the 459 sources with robust $\beta$ measurements (signal-to-noise ratio, S/N, $\geq$ 5) in three redshift intervals (z = 6-8, 8-10, 10-12). Bin uncertainties are estimated via bootstrap resampling: for each of 1000 iterations we resample with replacement, perturb redshifts and UV slopes within their uncertainties, re-bin the mock sample, and compute the standard deviation of each bin. The final bin errors are taken as the median of these standard deviations across all iterations. These median bins are plotted as red diamonds in Figure \ref{fig:beta-z} and summarized in Table \ref{tab:betaz_bins}.

The median UV slopes become slightly bluer with increasing redshift, but remain consistent at the 1-$\sigma$ level. This consistency remains when repeating the binning procedure with the full sample or with a stricter S/N $\geq 10$ cut. A Kendall's $\tau$ test yields $\tau$ = $-0.056$ (p = 0.05), indicating no statistically significant decrease with redshift within our sample. However, we highlight the lack of red galaxies at high redshift: only 11$\pm$5.1\% of galaxies at z $\geq$ 9 have $\beta > -2.0$, compared to 27$\pm$2.3\% as z $\leq$ 7. 

We fit a linear redshift relation to all S/N $\geq$ 5 sources (blue diamonds in Figure \ref{fig:beta-z}), finding d$\beta$/dz = $-0.031\pm0.016$. This is in agreement with the fainter subsample in \citet{Topping2024} (d$\beta$/dz = $-0.008\pm0.017$, $\langle$\muv$\rangle \sim -18$), but much shallower than found by \citet{Cullen2024} (d$\beta$/dz = $-0.28^{+0.05}_{-0.05}$, $\langle$\muv$\rangle \sim -19$). Differences in the redshift evolution of $\beta$ have been linked to a dependence on \muv, with brighter samples showing a steeper evolution \citep{Robertson2022, Topping2024}. The weak evolution measured here may reflect the faint average magnitude of the GLIMPSE sample ($\langle \rm{M_{UV}} \rangle \sim -16.5$). To explore this further, we split the GLIMPSE sample into three $M_{\rm UV}$ bins: \muv\ $<-17$, $-17 \leq$ \muv\ $ \leq -15$, and \muv\ $> -15$. At fixed \muv, we find no significant difference in the redshift dependence of $\beta$ in each sub-sample relative to the full sample.

%--Fig3: Beta-z

%--Tab2: median bins for beta-z
\begin{table}[h]
    \centering
    \caption{Median redshift, UV slope, M$_{\rm UV}$ and number of robust sources in each of the three redshift bins shown in Figure~\ref{fig:beta-z}.}
    \label{tab:betaz_bins}
    \begin{tabular}{lcccc}
        \hline
        Redshift bin & z$_{\rm phot}$ & $\beta$ & M$_{\rm UV}$ & N \\
        \hline
        $6 \leq z < 8$   & $6.67\pm0.23$ & $-2.28\pm0.22$ & $-16.8$ & 379 \\
        $8 \leq z < 10$  & $8.54\pm0.23$ & $-2.46\pm0.24$ & $-16.4$ & 63 \\
        $10 \leq z < 12$ & $11.1\pm0.28$ & $-2.52\pm0.24$ & $-16.7$ & 14 \\
        \hline
    \end{tabular}
\end{table}

\subsection{$\beta-M_{UV}$ Relation}\label{sec:beta-muv}

\begin{figure}[h]
    \centering
    \includegraphics[width=\linewidth]{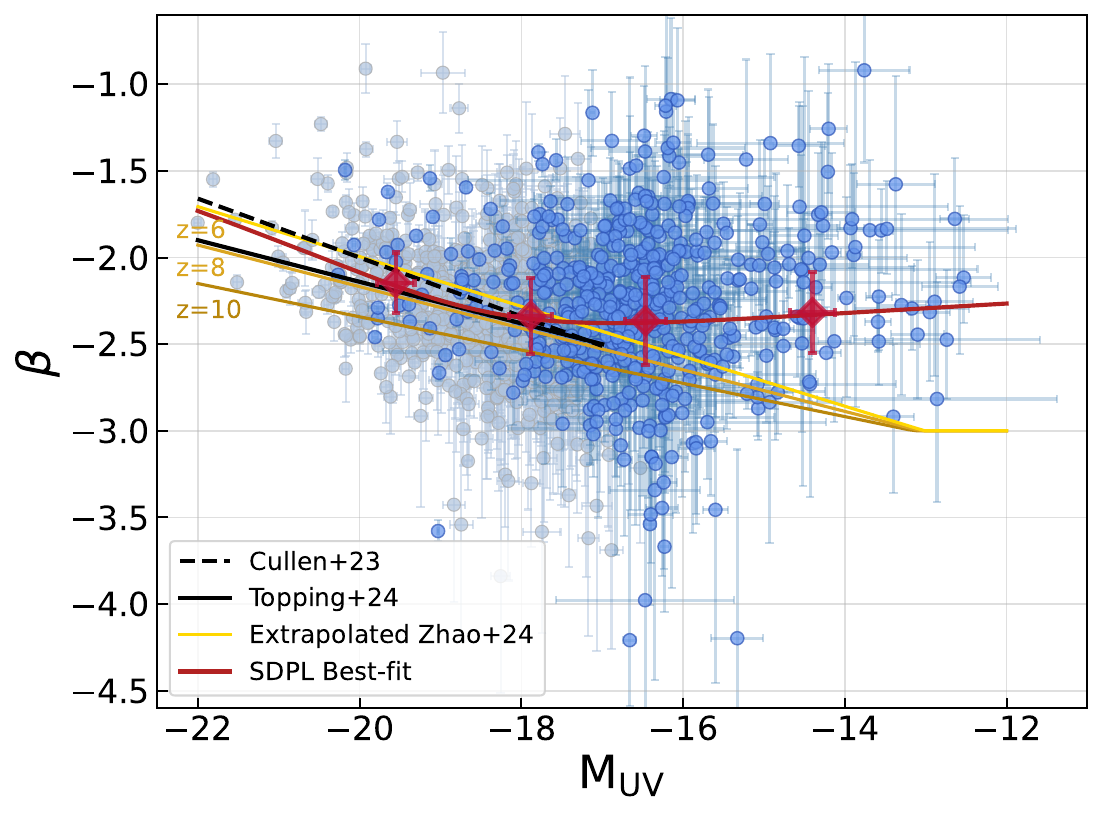}
    \caption{UV continuum slope versus \muv, showcasing the redder UV slopes of the faintest galaxies. We plot the full GLIMPSE sample (blue circles) as well as the JADES photometric sample of \citet{Endsley2024} (gray circles). Red diamonds are median bins for robust UV slopes with S/N $\geq$ 5 and the red curve is the best-fit of all robust points. Best-fit relations from \citet{Cullen2023,Topping2024} are plotted as solid and dashed black lines, respectively. Yellow lines show the combined \hst\ + JWST sample of \citet{Zhao2024}, extrapolated to fainter magnitudes for redshifts z=6,8,10.}
    \label{fig:beta-muv}
\end{figure}

Due to the intrinsic depth reached by GLIMPSE, we measure the UV slopes of 136 galaxies with \muv\ $\geq -16$, 87 of which have S/N $\geq$ 5. In Figure \ref{fig:beta-muv}, we show the dependence of $\beta$ on \muv\ for our full sample (blue points). To directly compare to brighter galaxies, we also calculate the UV slopes of the JADES photometric sample of \citet{Endsley2024} (gray points) following the methods outlined in Sections \ref{sec:uv_slopes}. The UV slopes of this JADES sample are independently calculated and fully analyzed in \citet{Topping2024}; we include them here for comparison. Following the procedure described in Section \ref{sec:beta-z}, we bin all robust GLIMPSE+JADES UV slopes into equal-width \muv\ intervals, with median values reported in Table \ref{tab:betamuv_bins}. We note the large variance in UV slopes across all magnitudes, particularly near the detection threshold for unlensed galaxies ($M_{\rm UV} \simeq -17$). 

Consistent with previous studies \citep[e.g.][]{Cullen2023, Topping2024, Austin2024}, we find that UV slopes initially become bluer towards fainter magnitudes over the range $M_{\rm UV} \in [-22, -17]$. In this range, our median bins agree with the best-fit trends of \citet{Cullen2023, Topping2024} while remaining consistent with one another within their uncertainties. Furthermore, \citet{Zhao2024} fit a relation for the average UV slope as a function of both redshift and magnitude from the z=4-8 \hst\ sample of \citet{Bouwens2014} and the z=6-12 JWST sample of \citet{Topping2024}, both which reach down to \muv\ $\simeq -17$. We over-plot this decreasing relation for three redshifts as the yellow curves in Figure \ref{fig:beta-muv}, extrapolating to fainter magnitudes and capping $\beta$ at $-3.0$. 

Notably, the decreasing trend of $\beta$ with magnitude does not extend to fainter galaxies now observable with GLIMPSE. We find that \textit{the faintest galaxies show UV slopes no bluer than those of brighter galaxies}. Indeed, our faintest bin at \muv\ $= -14.4$ is consistent with all brighter median bins and lies $\sim 2\sigma$ above the extrapolated trend ($\langle \beta \rangle = -2.31$ compared to predicted $\beta \simeq -2.9$). Repeating the binning procedure including lower signal points drives the faintest bin even redder ($\langle \beta \rangle = -2.08$). 

To test whether the $\beta-$\muv\ is better described by a simple trend or two distinct regimes, we fit both a single power-law and a smooth double power-law (SDPL) to the robust S/N $\geq$ 5 GLIMPSE+JADES sample. We find the data is best represented by the smooth double-power law (red curve, Figure \ref{fig:beta-muv}) given by:
\begin{equation}
    \beta(M) = \frac{\beta_0}{\left[10^{-0.4(M-M^*),\alpha_{1} * n} + 10^{-0.4(M-M^*),\alpha_{2} * n}\right]^{1/n}},
\end{equation}
where $\beta_0 =-2.45 \pm 0.14$ is the normalization, $M^* = -18.5 \pm 1.00$ is the characteristic magnitude separating the bright- and faint-end behavior, $n = 11.5\pm16.6$ sets the sharpness of this transition, and $\alpha_{1} = 0.11\pm0.06$ and $\alpha_{2} = -0.01\pm0.02$ are the bright- and faint-end slopes, respectively. These values indicate that while bright galaxies get progressively bluer as luminosity decreases, galaxies fainter than \muv\ $\sim -18$ either stay as blue as their brighter counterparts or get slightly redder. The SDPL fits the S/N $\geq$ 5 data slightly better compared to a single power-law ($\chi^2_{\rm red, SDPL} \sim 8.83$, $\chi^2_{\rm red, PL} \sim 10.5$) with a $\Delta$BIC = 21 strongly favoring the SDPL fit. Hence, the faintest galaxies have UV slopes possibly redder, but at least no bluer than those of brighter galaxies. 

Furthermore, this faint galaxy population is expected to have an average UV slope biased toward bluer values. First introduced by \citet{Dunlop2012}, this \lq{}$\beta$ bias' results when the filter red-ward of the Lyman break has its flux boosted into the detection threshold by photometric noise or scatter, resulting in an artificially bluer UV slope. This bias is most prominent for galaxies near the detection limit in a Lyman break selected sample where the filters used in source selection are also used in calculating $\beta$ \citep{Bouwens2012, Austin2024, Asada2025}. Galaxies scattered to bluer values will be preferentially included, biasing the median UV slope to bluer values. Since the majority of our faint galaxies are detected via the Lyman-break selection (92\%), it is likely that the median UV slope of our faintest galaxies are even redder than observed, further reinforcing that the faintest galaxies do not have the bluest UV slopes.

%--Fig4: Beta-Muv

%--Tab3: Median bins for muv-z
\begin{table}[h]
    \centering
    \caption{Median \muv, UV slope, redshift, and number of robust GLIMPSE+JADES sources in each \muv\ bin shown in Figure~\ref{fig:beta-muv}.}
    \label{tab:betamuv_bins}
    \begin{tabular}{lcccc}
        \hline
        M$_{\rm UV}$ bin & M$_{\rm UV}$ & $\beta$ & z$_{\rm phot}$ & N \\
        \hline
        $[-21, -19]$   & $-19.6\pm0.23$ & $-2.14\pm0.17$ & $6.68$ & 224 \\
        $[-19, -17]$  & $-17.9\pm0.27$ & $-2.34\pm0.22$ & $6.87$ & 654 \\
        $[-17, -15]$ & $-16.5\pm0.25$ & $-2.37\pm0.25$ & $6.77$ & 264 \\
        $[-15, -13]$ & $-14.4\pm0.28$ & $-2.31\pm0.23$ & $7.17$ & 28 \\
        \hline
    \end{tabular}
\end{table}

%%%%%%%%%%%%%%%%%%%%%%%%%%%%%%%%%%%%%%%%%%%%%%%%%%%%%%%%%%%%%%%%%%%%%%%%%%%%%%%%%%%%%%%%%%%%%%%%%%%%%%%%%%%%%%%%%%%%%%%%%%%%%%%
\section{Discussion} \label{sec:disc}
%%%%%%%%%%%%%%%%%%%%%%%%%%%%%%%%%%%%%%%%%%%%%%%%%%%%%%%%%%%%%%%%%%%%%%%%%%%%%%%%%%%%%%%%%%%%%%%%%%%%%%%%%%%%%%%%%%%%%%%%%%%%%%%

We have presented the UV continuum slopes of GLIMPSE galaxies, highlighting the surprisingly redder $\beta$ values of the faintest galaxies. Here we look more in-depth at both the variety of faint galaxies and galaxies with $\beta \leq -2.8$, contextualizing their UV slopes through their observed emission-line properties. We then justify and comment on the implied LyC escape fractions from these UV slopes, validating our findings by comparing to MEGATRON simulations of faint galaxies \citep{Katz2025}. Finally, we discuss what the GLIMPSE observations of faint galaxies imply for the timing and drivers of cosmic reionization.  

\subsection{A Diverse Population of Faint Galaxies}\label{sec:faint}
%%%%%%%%%%%%%%%%%%%%%%%%%%%%%%%%%%%%%%%%%%%%%%%%%%%%%%%%%%%%%%%%%%%%%%%%%%%%%%%%%%%%%%%%%%%%%%%%%%%%%%%%%%%%%%%%%%%%%%%%%%%%%%%

%--Fig7: Grid of faint, red gals
\begin{figure*}
    \centering
    \includegraphics[width=\textwidth]{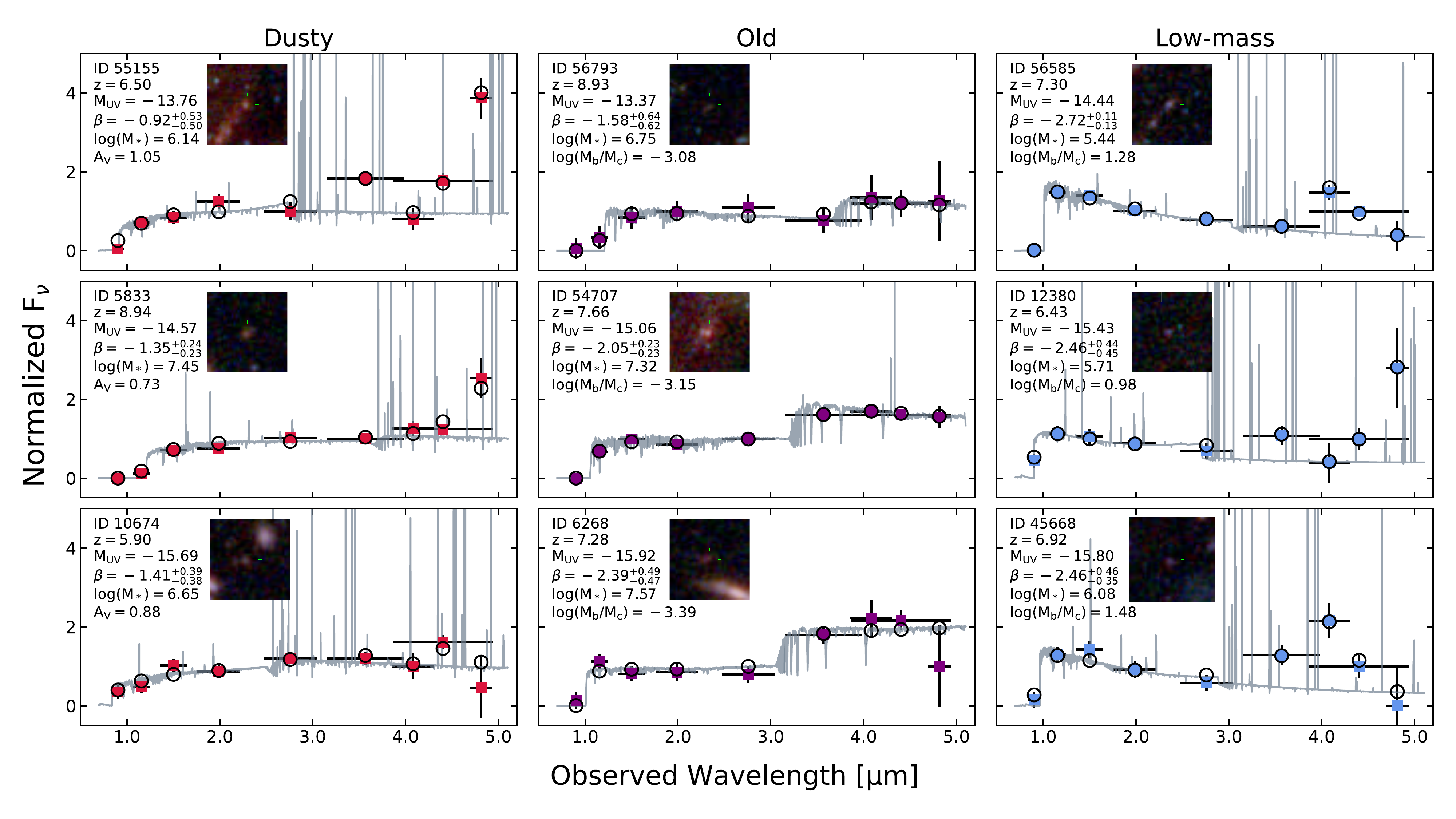}
    \caption{Notable faint (\muv\ $> -16$) galaxies observed with GLIMPSE, selected to emphasize the diversity of this subsample. Each column showcases galaxies whose faint magnitudes are primarily driven by dust attenuation (left), older stellar populations (middle), and low stellar masses (right). For each galaxy, we compare the observed photometry (squares) to \textsc{Bagpipes} model photometry (black circles) and show the best-fit SED as gray curves. We list A$_{\rm V}$ and intrinsic total stellar mass from the maximum-likelihood model. We also report the logarithmic ratio of the stellar mass formed in the recent burst to that formed in the older, constant star-formation component (M$_{\rm b}$/M$_{\rm c}$), indicating the relative mass-fraction of the stellar population.}
    \label{fig:faint}
\end{figure*}

The faintest galaxies are expected to have minimal dust, young ages, and low metallicity. These properties imply faint galaxies should have bluer UV slopes than brighter galaxies, as observed in the range $-20 < $ \muv\ $< -17$ for z $>$ 6 with JWST \citep[e.g.][]{Cullen2023, Topping2024} and z $<$ 8 with HST \citep{Rogers2013,Bouwens2014}. However, these properties have yet to be observationally verified for populations of galaxies with \muv\ $\gtrsim -16$. By leveraging strong gravitational lensing and ultra-deep imaging with JWST, the GLIMPSE survey provides the first statistical view of the faintest galaxies in the early universe.

We detect 136 galaxies with \muv\ $> -16$ and find a diverse population of faint galaxies. The UV slopes of faint galaxies span a wide range, with a dispersion of $\sigma_{\rm \beta} = 0.45$ for median $\beta = -2.27$ and \muv $ = -15.1$. We also observe a broad range of rest-optical emission line strengths, as clearly emphasized in Figure \ref{fig:faint} by the SEDs where the 410M or 480M filters cover [\ion{O}{3}]+H$\beta$ and H$\alpha$ emission. Furthermore, the maximum-likelihood \textsc{Bagpipes} models for the faint sub-sample predict an extreme range of dust attenuation (A$_{\rm V} \in$ [0, 1.05]) and intrinsic total stellar mass (log(M$_{\star}) \in$ [4.4, 8.1]). The large variance in observed properties suggests that galaxies appear faint for multiple reasons: dust attenuation, old stellar populations, and/or low stellar masses. 

The SEDs in Figure \ref{fig:faint} highlight the most extreme examples of galaxies whose faint magnitudes are primarily driven by each of these three mechanisms. The left column shows faint galaxies with red UV slopes ($\beta \sim -1.3$) and strong rest-optical emission, representative of dusty galaxies undergoing strong star-formation. The middle column features faint galaxies with moderately blue UV slopes ($\beta \sim -2.2$), minimal line-emission, and prominent Balmer breaks, indicative of older stellar populations ($>$ 10 Myr) with recently quenched star formation. \textsc{Bagpipes} models predict their stellar populations were primarily formed in the older, constant star-formation component rather than the younger burst component, resulting in low burst-to-constant mass fractions (M$_{\rm b}$/M$_{\rm c}$). Lastly, and most commonly expected, the right column shows faint galaxies with blue UV slopes ($\beta \sim -2.5$) and strong line-emission, requiring low stellar masses to explain their faint magnitudes, as predicted by \textsc{Bagpipes}.

Of course, most of our faint galaxies (\muv\ $> -16$) are influenced by more than one of these factors (e.g. both old and low-mass), but we can broadly separate our faint galaxy sample into these three populations. Around 30\% of our faint sub-sample have UV slopes redder than $-2.0$ (A$_{\rm V} \gtrsim 0.1$), suggesting measurable dust can be present in even the faintest systems at z $>$ 6. Around 15\% have predicted burst-to-constant mass fractions M$_{\rm b}$/M$_{\rm c} < 0.1$ from \textsc{Bagpipes}, indicating the bulk of their stellar mass formed in older star formation episodes rather than recent bursts. The majority of our observed faint galaxies are low mass, with $\sim 50\%$ of the sub-sample having log(\Mstar/\Msun) $< 6$. 

Although we see this diversity in galaxy properties, the scatter in UV slopes goes towards redder rather than bluer values. Only 26\% of galaxies with \muv\ $> -16$ have UV slopes bluer than $-2.5$, and no galaxy fainter than \muv\ $= -15$ has $\beta \leq -3.0$. Thus, we observe comparable numbers of dust-free and dusty faint galaxies. In comparison, slightly brighter galaxies with \muv\ $\in [-19,-17]$ show a much lower fraction of red UV slopes (16\% with $\beta > -2.0$) and a slightly higher fraction of blue slopes (30\% with $\beta < -2.5$).

Overall, we find that faint galaxies are not a monolithic population, but rather encompass a range of stellar masses, dust content, and star-formation histories. These GLIMPSE observations suggest that early dust production and bursty star formation histories complicate the interpretation of faint galaxies. We find that a single UV luminosity can translate to a diversity of galaxy properties, making it difficult to directly map the UVLF onto the halo mass function. The observed diversity in the faint-end UVLF agrees with past work that suggests stochastic star formation drives the over-abundance of UV-bright galaxies observed by JWST in the early universe \citep{Mason2023, Kravtsov2024, Gelli2024}. Low-mass galaxies under-going a strong burst of star-formation appear brighter than galaxies of similar mass undergoing more modest star formation. This scatters galaxies at fixed halo mass into different observed luminosities. The GLIMPSE observations suggest a similar scattering occurs due to dust, where galaxies in the early universe have a range of dust attenuation at fixed halo mass \citet{Ferrara2025}. Dustier galaxies will appear fainter than dust-free galaxies at the same halo mass. Therefore, both star formation history and dust production shift early galaxies to fainter or brighter magnitudes than theoretically predicted based on their halo masses.

\subsection{Extremely Blue Objects}\label{sec:extblue}
%%%%%%%%%%%%%%%%%%%%%%%%%%%%%%%%%%%%%%%%%%%%%%%%%%%%%%%%%%%%%%%%%%%%%%%%%%%%%%%%%%%%%%%%%%%%%%%%%%%%%%%%%%%%%%%%%%%%%%%%%%%%%%%

%--Fig5: SED fits w varying fesc
\begin{figure*}
    \centering
    \includegraphics[width=\linewidth]{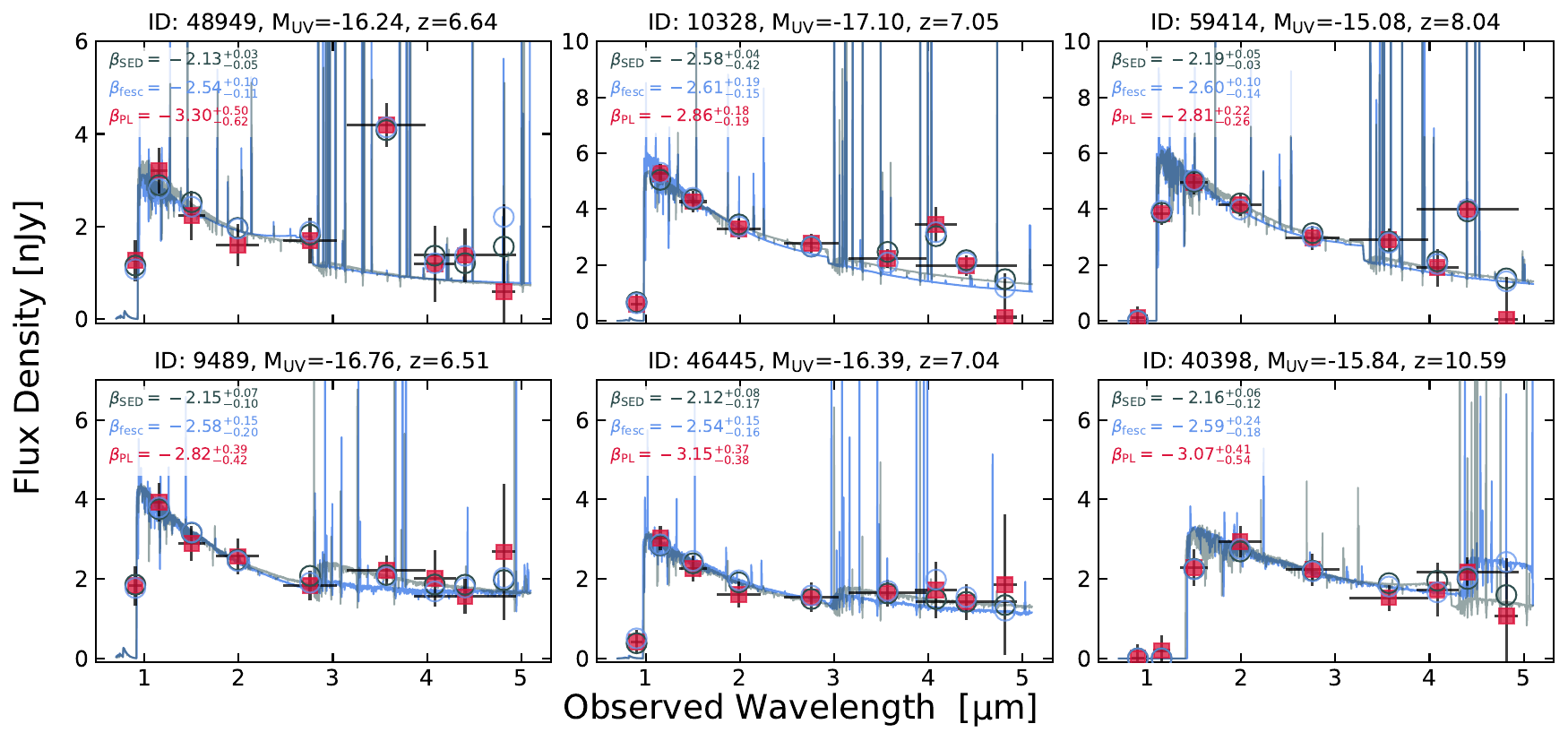}
    \caption{Sample of galaxies with the most robust extremely blue UV slopes ($\beta \leq -2.8$) which would require no dust attenuation and nebular continuum. The observed photometry is plotted as red squares. Best-fit SEDs and model photometry from standard \textsc{Bagpipes} model with \fesc=0 are shown in gray. Adapted \citet{Giovinazzo2025} models with varying \fesc\ are shown in blue. We list the UV slope calculated directly from photometry ($\beta_{\rm PL}$), the \fesc=0 SED ($\beta_{\rm SED}$), and the varying \fesc\ SED ($\beta_{\rm fesc}$).}
    \label{fig:veryblue}
\end{figure*}

Galaxies with extremely blue UV continuum slopes ($\beta \leq -2.8$) are of particular interest, as standard population synthesis models with ionization-bounded conditions struggle to reproduce these very blue UV slopes \citep{Raiter2010}. Since $\beta$ is reddened by nebular continuum and dust attenuation, an observed UV continuum bluer than $\beta \sim -2.8$ cannot be dominated by nebular continuum and must have no dust. Extremely blue UV slopes have been suggested to be caused by high LyC escape fractions or exotic stellar populations, but often result from photometric uncertainties (see Section \ref{sec:beta-muv}).
 
We find 52 galaxies with $\beta \leq -2.8$ and assess whether their extremely blue UV slopes are robust. Figure \ref{fig:veryblue} shows the six most robust extremely blue candidates, all of which have robust $\beta$ measurements with S/N $\geq$ 5 and observed photometry consistent within 1-$\sigma$ with all model fluxes predicted by the adapted \textsc{Bagpipes} model of \citet{Giovinazzo2025}. For the majority of galaxies with $\beta \leq -2.8$, the UV slope uncertainties are large, with a median $\sigma_{\beta}$ = 0.49, much larger than that of the full sample ($\sigma_{\beta}$ = 0.29). From a visual inspection of each source, we find most extremely blue galaxies exhibit noisy SEDs, often with an unusually flat color between the F200W and F277W filters, suggesting F200W is under-measuring the flux and responsible for driving the UV slope artificially bluer, as shown in the upper left panel of Figure \ref{fig:veryblue}. Overall, we find that the majority of galaxies with extremely blue $\beta$ values lack convincing evidence for such extreme slopes and can be explained by photometric uncertainty.

Furthermore, many of these galaxies have observed magnitudes near the 5-$\sigma$ broadband detection threshold ($\langle$M$_{\rm obs}\rangle$ = 30.07 mag, $\rm{M}_{\rm obs,\ thresh} = 30.8\ \rm{mag}$), where extreme UV slopes can bias source selection (as discussed in Section \ref{sec:beta-muv}). As most of these galaxies are also weakly magnified by gravitational lensing, this bias accounts for the increased variance in observed UV slopes at the corresponding absolute magnitude of \muv\ $\simeq -17$ in Figure \ref{fig:beta-muv}. Even with the increased depth of GLIMPSE, we do not find more convincing extreme UV slope candidates than shallower surveys \citep[e.g.][]{Cullen2024, Topping2024}.

Notably, our extremely blue UV slopes are not from the faintest galaxies: only 13 (25\%) sources with $\beta \leq -2.8$ have \muv\ $> -16$. If these blue UV slopes were indicative of Population III objects or extreme low metallicity, we would expect to find them at the faintest magnitudes.

Although the majority of galaxies with extremely blue UV slopes have large uncertainties, we do identify individual galaxies with compelling evidence for very blue $\beta$ values driven by high LyC escape fractions. In the high \fesc\ scenario, there cannot be dust and ionizing photons are not absorbed by the hydrogen in the galaxy to create the nebular continuum, both of which strongly redden $\beta$. If our extremely blue UV slopes are indeed caused by high \fesc, they should have reduced rest-optical line-emission as shown by galaxies in the bottom row of Figure \ref{fig:veryblue}.

To further explore whether any extremely blue UV slopes in our sample indicate high LyC escape fractions, we analyze the \textsc{Bagpipes} picket-fence model fits of \citet{Giovinazzo2025}. Introducing \fesc\ as a free parameter allows the \textsc{Bagpipes} model to reproduce an SED with both $\beta \leq -2.8$ and weak emission lines. As highlighted in the lower panel of Figure \ref{fig:betabeta}, we find that the varying \fesc\ model frequently results in an SED with a bluer slope, finding an average decrease of $\Delta \beta = -0.25$ for galaxies with extremely blue UV slopes. We compare the maximum likelihood SEDs for the zero \fesc\ model (gray curve) to the varying \fesc\ model (blue curve) for each galaxy in Figure \ref{fig:veryblue}. Allowing \fesc\ to vary only slightly improves the model fits, decreasing the median $\chi^2_{\nu}$ from 0.67 to 0.64. This limited improvement while driving the UV slopes bluer is likely due to the age-\fesc\ degeneracy: galaxies with extremely blue UV slopes and weak emission lines could be explained as either high \fesc\ or a recently quenched B-Star dominated stellar population \citep{Giovinazzo2025}. With only photometry, it is difficult to disentangle these options.

Lastly, we have many galaxies with very blue, but not extreme, UV slopes of $\beta \sim -2.6$ that could have non-negligible LyC escape fractions. The JWST Director's Discretionary Time program GLIMPSE-\textit{D} (PID 9223; PIs: S. Fujimoto \& R.P. Naidu) obtained deep ($\sim$ 30.4 hr) NIRSpec/MSA G395M observation of the Abell S1063 field, as detailed in \citet{Fujimoto2025}. The spectra cover rest-frame H$\alpha$ for $\sim 60$ galaxies in our sample, most of which have no continuum emission detected. 

\begin{figure*}
    \centering
    \includegraphics[width=\linewidth]{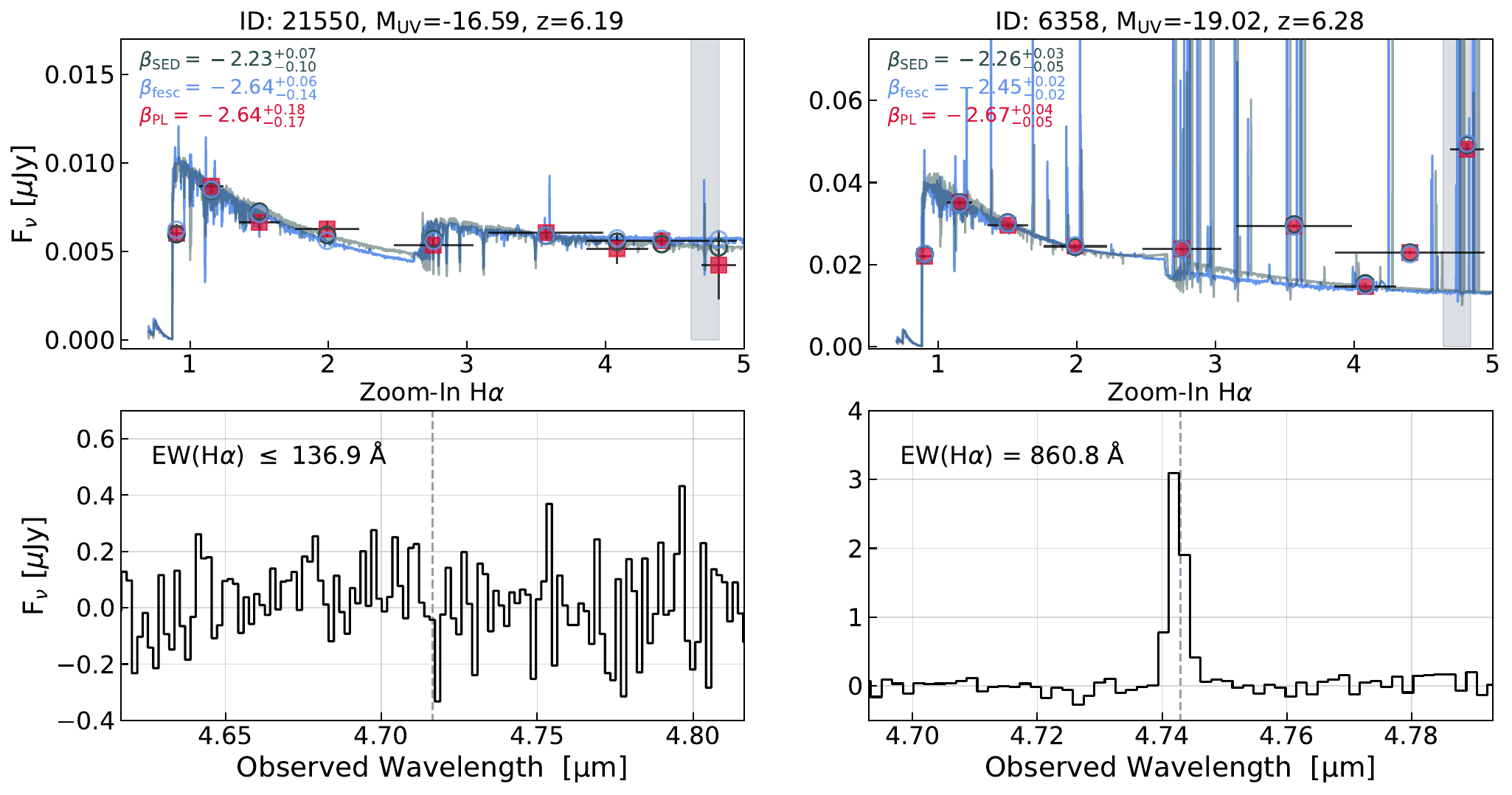}
    \caption{Two galaxies with blue UV slopes ($\beta \sim -2.6$) for which we have NIRSpec/MSA G395M coverage. In the top row, we show their observed photometry (red squares) and best-fit SEDs, same as in Figure \ref{fig:veryblue}. In the bottom row, we show the corresponding observed spectra near H$\alpha$ (gray shaded region), and report the equivalent width or upper limit, if undetected. }
    \label{fig:ddtspec}
\end{figure*}

We highlight two galaxies observed by this program in Figure \ref{fig:ddtspec}, showing both the GLIMPSE photometry and best-fit SEDs (upper panels) and zooming in on their NIRSpec H$\alpha$ emission. Both galaxies have similarly blue UV slopes ($\beta \sim -2.65$) but different rest-optical line-emission strengths. In the left panel, ID 21550 has no detected H$\alpha$ emission, with a rest-frame upper limit of EW(H$\alpha$) $\leq$ 137 Å. This low H$\alpha$ emission suggests either a very high escape fraction or a $\sim$10 Myr stellar population. We tested the possibility of a low-redshift solution by forcing a z$_{\rm phot} < 5$ \textsc{Bagpipes} model, which predicts a detectable Pa$\alpha$ emission line (observed-frame EW(Pa$\alpha$) $\simeq$ 600 $\AA$) that we do not observe in the spectra. In comparison, ID 6358 has moderately strong H$\alpha$ emission (EW(H$\alpha$) = 860 Å), as shown in the right panel, requiring recent star-formation while non-negligible LyC escape remains feasible. 

These spectroscopic observations highlight the diversity of very blue GLIMPSE galaxies. While there is not a large population of extremely blue galaxies in GLIMPSE, there are a handful of very blue galaxies with weak nebular emission lines. If these galaxies are still producing ionizing photons, they must have significant \fesc\ and would be important for cosmic reionization. See Asada et al.\ (in prep.) for spectroscopy of such a GLIMPSE-\textit{D} source. 

%--Fig6: DDT spec

\subsection{Escape Fractions}
%%%%%%%%%%%%%%%%%%%%%%%%%%%%%%%%%%%%%%%%%%%%%%%%%%%%%%%%%%%%%%%%%%%%%%%%%%%%%%%%%%%%%%%%%%%%%%%%%%%%%%%%%%%%%%%%%%%%%%%%%%%%%%%

%--Fig7: Beta-Halpa EW; Beta-Xi_ion
\begin{figure*}
    \centering
    \includegraphics[width=\textwidth]{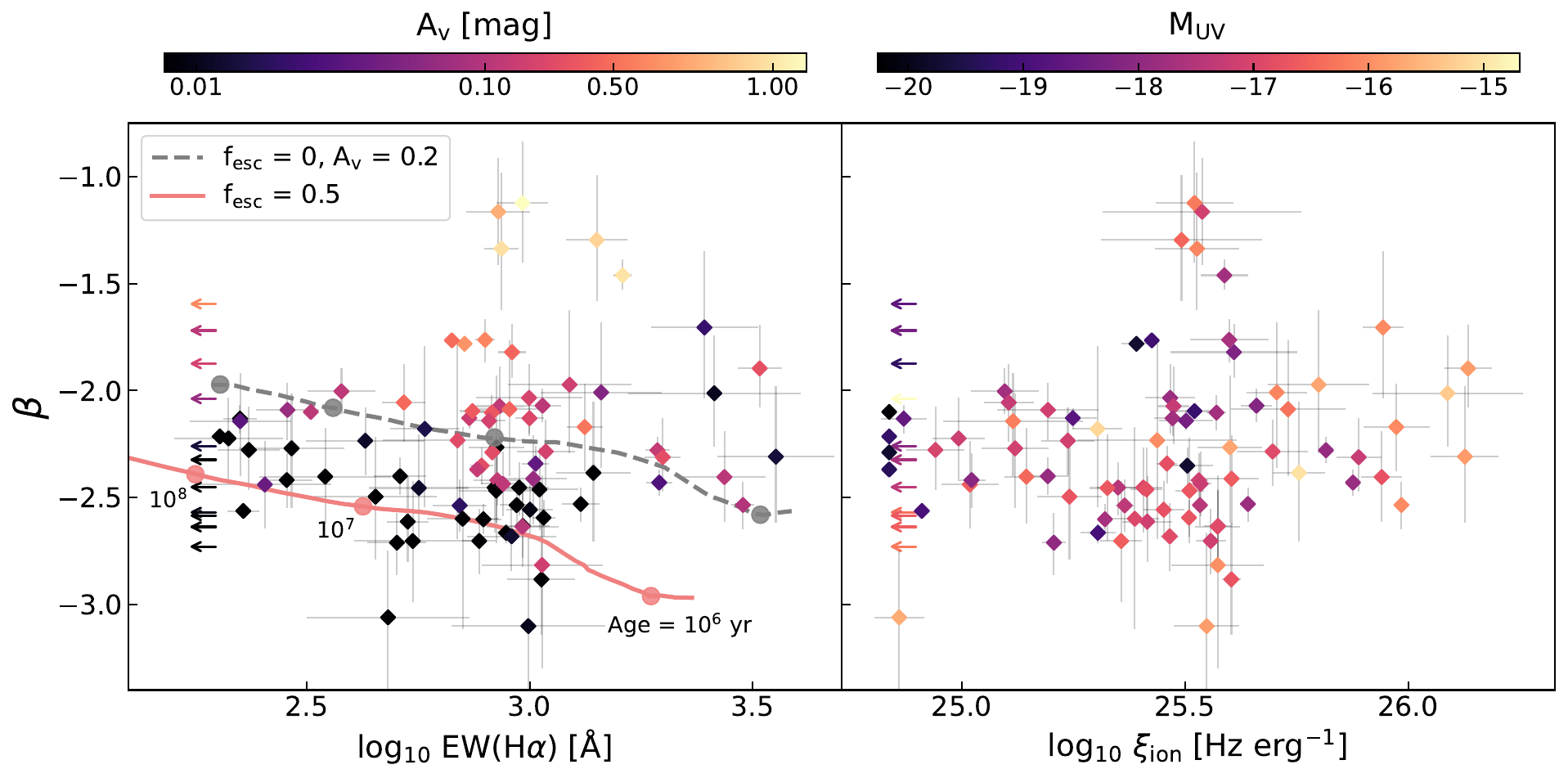}
    \caption{The relation between UV slope, rest-frame H$\alpha$ EW, $\xi_{\rm {ion}}$, A$_{\rm V}$, age, \fesc, and \muv\ for the z$\sim$6.3 photometrically selected sample where H$\alpha$ falls into the F480M filter. 
    \textbf{Left:} $\beta$ as a function of H$\alpha$ EW, coloring each point according to A$_{\rm V}$. We show two stellar population synthesis models from \citet{Zackrisson2013}: the \fesc=0, A$_{\rm V}$=0.2 model (dashed gray curve) for which points above require zero LyC escape with increasing dust attenuation and the \fesc=0.5 model with no dust (pink curve) for which points below likely have nonzero \fesc.
    \textbf{Right:} $\beta$ as a function of $\xi_{\rm {ion}}$, where points are colored according to \muv. We do not observe a significant trend between $\beta$ and $\xi_{\rm {ion}}$
    }
    \label{fig:xiion}
\end{figure*}

Constraining the escape fractions of galaxies in the EoR is notoriously challenging, yet essential for identifying the sources responsible for reionization. At z $>$ 4 direct detections of LyC are improbable due to absorption from neutral hydrogen in the IGM \citep{Inoue2014}, so \fesc\ is often estimated through indirect tracers calibrated in the local universe. A current gold standard for estimating \fesc\ is a multivariate predictor combining information on dust attenuation, nebular properties, and ionization state \citep{Flury2022, SaldanaLopez2022, Jaskot2024, Mascia2024}. Since the UV continuum slope traces both the dust and gas content of a galaxy, it is a powerful single variable predictor of \fesc\ \citep{Chisholm2022}, especially for populations of faint galaxies at high redshift where additional indirect indicators are observationally infeasible. 

To contextualize whether our UV slopes are indeed correlated with \fesc, we analyze the relations between $\beta$, A$_{\rm V}$, H$\alpha$ emission, and \muv\ for the photometrically selected z $\sim$ 6.3 sample of Chisholm et al.\ (in prep.). The rest-frame H$\alpha$ equivalent widths (EW) are calculated from the F480M photometric excess and the dust-corrected $\xi_{\rm {ion}}$ is measured from the H$\alpha$ luminosity and SED fits of Chisholm et al.\ (in prep.), which assume an SMC dust curve. We note these values are lower limits because any escaping LyC would reduce H$\alpha$ emission, leading to an underestimation of the intrinsic $\xi_{\rm {ion}}$.

The left panel of Figure \ref{fig:xiion} shows EW(H$\alpha$) versus $\beta$, with points colored by their best-fit A$_{\rm V}$ from the standard \textsc{Bagpipes} model. The strong correlation between $\beta$ and A$_{\rm V}$ emphasizes that the UV slope primarily traces dust attenuation. We overlay the constant star-formation rate stellar population synthesis models with \fesc=0 and \fesc=0.5 from \citet{Zackrisson2013}, adding dust attenuation to the zero \fesc\ model. As shown by the pink curve, the UV slope evolves minimally with age, reddening by $\Delta \beta \sim 0.5$ over 100 Myr. The right panel of Figure \ref{fig:xiion} shows $\xi_{\rm {ion}}$ versus $\beta$, now colored according to \muv.

We do not find a simple correlation between $\beta$ and H$\alpha$ or $\xi_{\rm {ion}}$; instead, the trend is complicated by stellar population age and \fesc. Galaxies with the largest H$\alpha$ EW (log$_{\rm 10}$EW(H$\alpha$) $\sim 3.5$ Å) have moderately red UV slopes ($\beta \sim -2.0$), requiring low escape fractions because ionizing photons must be absorbed to power Balmer emission and nebular continuum. These galaxies lie above the \fesc=0 model in the upper-right corner of the left panel of Figure \ref{fig:xiion} and tend to be slightly fainter than the full sample. In contrast, galaxies with the bluest UV slopes do not exhibit large $\xi_{\rm {ion}}$ values. In particular, no galaxy with $\beta \leq -2.5$ has H$\alpha$ equivalent width above 1075 Å, placing them below the \fesc=0.5 model and suggesting partial LyC leakage. Overall, we see that galaxies with red UV slopes have signatures of low \fesc\ from their strong nebular emission lines, while galaxies with blue UV slopes have rest-frame optical emission line properties that are consistent with non-zero \fesc. This behavior contrasts with the modest correlation between increasing $\xi_{\rm {ion}}$ and bluer UV slopes reported in previous studies \citep{Bouwens2016,Shivaei2018,Prieto2023,Begley2025}. By probing ultra-faint galaxies we discover galaxies that are older and have higher dust content (see Section \ref{sec:faint}), therefore we find a flatter $\beta -\xi_{\rm {ion}}$ relation with significant scatter.

As this subsample is photo-z selected, it is biased towards galaxies with strong emission. Chisholm et al.\ (in prep.) demonstrate that our faintest galaxies are preferentially observed during a burst of star-formation with elevated $\xi_{\rm {ion}}$. This suggests we may be missing a subset of faint galaxies with no recent star-formation that would occupy the leftmost region of Figure \ref{fig:xiion}. However, their older stellar population ages and the overall red UV slopes of the faint galaxies in the Lyman-break sample imply this population is unlikely to be significantly contributing to reionization.

In general, this multivariate analysis supports the interpretation that UV continuum slopes trace LyC escape at high redshift and faint magnitudes. Thus, we estimate the escape fraction for all galaxies in our sample using the $\beta-$\fesc\ relation derived from local galaxies by \citet{Chisholm2022}, capping \fesc\ at unity. As there is significant scatter in the empirical relation, this approach is best suited for estimating population-level trends rather than accurate measurements of \fesc\ for individual galaxies. To estimate uncertainties, we perturb $\beta$ and the $\beta$-\fesc\ fit parameters within their respective errors and recompute \fesc\ for 1000 iterations, adopting the inner 68th percentile of the resulting \fesc\ distribution as the uncertainty.

\begin{figure}[h]
    \centering
    \includegraphics[width=\linewidth]{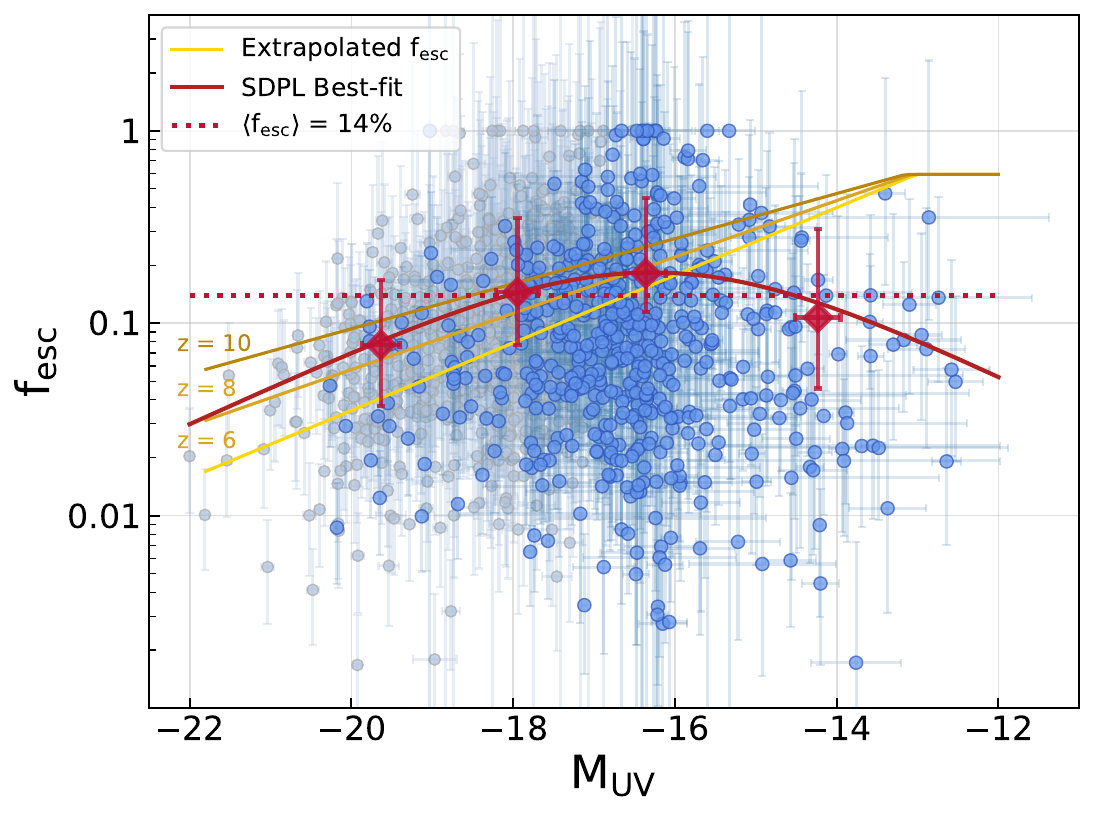}
    \caption{The modest escape fractions of the faintest galaxies now observable with GLIMPSE (blue circles) alongside the brighter JADES sample (gray circles). Red diamonds are mean \fesc\ for equal-width \muv\ bins of robust galaxies with S/N $\geq$ 5 and the solid red curve is best-fitting relation to all robust galaxies. We compare to the commonly used prediction from applying the local $\beta$-\fesc\ relation on the UV slopes of \citet{Zhao2024}, extrapolated to fainter magnitudes at three redshifts (yellow curves).}
    \label{fig:fesc-muv}
\end{figure}

Figure \ref{fig:fesc-muv} shows \fesc\ as a function of absolute magnitude, again showing the faint galaxies detected by GLIMPSE (blue circles) alongside the brighter JADES photometric sample \citep[gray circles;][]{Endsley2024}. We calculate the mean \fesc\ in bins of \muv\ for robust galaxies with S/N $\geq$ 5 (red diamonds) following the procedure of Section \ref{sec:res}, now using the average instead of median. We list the bin values in Table \ref{tab:fescmuv_bins}. The large scatter in predicted \fesc\ at all magnitudes results in the large binned error values.

We find that the faintest galaxies (\muv\ $> -16$) do not have the highest escape fractions, but are, at best, consistent with brighter galaxies – as indicated by their UV continuum slopes. For comparison, we show the inferred evolution of \fesc\ with \muv\ by applying the $\beta-$\fesc\ relation to the \citet{Zhao2024} $\beta-$\muv\ trend extrapolated towards fainter galaxies (yellow curves). These extrapolations agree with the z $>$ 6 predictions of \citet{Chisholm2022} and predicted that the faint galaxies observed with GLIMPSE would have near unity escape fractions which require extremely blue UV slopes ($\beta \sim -3.0$). Instead, the extrapolation overpredicts by more than 1$\sigma$ the escape fraction of the majority of our faintest galaxies.

Once again, we fit all robust S/N $\geq$ 5 GLIMPSE+JADES sources with both a SDPL and a single power-law ($\chi^2_{\rm red, SDPL} \sim 22.2$, $\chi^2_{\rm red, PL} \sim 31.1$), finding the SDPL with the following parameters to be the best fit: $\beta_0 = 0.39 \pm 0.94$, $M^* = -16.2 \pm 2.2$, $n = 0.91 \pm 2.9$, $\alpha_{1} = 0.48 \pm 0.44$, and $\alpha_{2} = -0.51 \pm 2.01$. These values indicate the escape fraction peaks at $\sim 20\%$ for galaxies with \muv\ $\sim -16$ and declines sharply for fainter galaxies. Although the data is best-fit by a SDPL (solid red curve), we note that our mean \fesc\ bins are consistent with a constant $\sim 14\%$ escape fraction across all magnitudes (dashed red line), in agreement with findings of \citet{Giovinazzo2025} from SED-fitting and \citet{Mascia2024} from multivariate predictors but higher than the SED-inferred \fesc\ of \citet{Papovich2025}. 

This work provides the first population-level estimations for the expected escape fractions of galaxies with \muv\ $> -16$, although not without caveats. The observed, low-z $\beta-$\fesc\ relation shows roughly an order of magnitude scatter in \fesc\ at fixed UV slope, leading to large uncertainties on \fesc\ for individual galaxies. Nevertheless, the overall trend suggests that the faintest galaxies, in general, have escape fractions comparable to the full sample. In addition, we have assumed the $\beta-$\fesc\ relation measured from local galaxies with $\beta > -2.6$ extends to higher redshifts, fainter magnitudes, and bluer galaxies. Encouragingly, initial evidence shows this relation agrees with other multivariate estimates of \fesc\ at high redshift \citep{Gazagnes2025}. These \fesc\ predictions provide the first empirically constrained estimates of the escape fractions of the faintest galaxies, offering a direct look at their role in reionization. 

% --Fig8: fesc-Muv

%--Tab4: Mean bins for muv-fesc
\begin{table}[h]
    \centering
    \caption{Mean and median \fesc\ for the four S/N $\geq$ 5 \muv\ bins shown in Figure \ref{fig:fesc-muv}. Same magnitude bins as Table \ref{tab:betamuv_bins}.}
    \label{tab:fescmuv_bins}
    \setlength{\tabcolsep}{6pt}
    \begin{tabular}{cccc}
        \hline
        M$_{\rm UV}$ bin & M$_{\rm UV}$ & $\langle$\fesc $\rangle$ & Median \fesc \\
        \hline
        $-21 \leq M_{\rm UV} < -19$   & $-19.6\pm0.23$ & $0.08^{+0.09}_{-0.04}$ & $0.05^{+0.09}_{-0.04}$ \\
        $-19 \leq M_{\rm UV} < -17$  & $-17.9\pm0.27$ & $0.14^{+0.21}_{-0.07}$ & $0.09^{+0.21}_{-0.07}$\\
        $-17 \leq M_{\rm UV} < -15$ & $-16.5\pm0.25$ & $0.18^{+0.26}_{-0.07}$ & $0.10^{+0.27}_{-0.07}$\\
        $-15 \leq M_{\rm UV} < -13$ & $-14.4\pm0.28$ & $0.11^{+0.20}_{-0.06}$ & $0.09^{+0.19}_{-0.06}$\\
        \hline
    \end{tabular}
\end{table}

\subsection{Comparison of Faint Galaxies to Simulations}
%%%%%%%%%%%%%%%%%%%%%%%%%%%%%%%%%%%%%%%%%%%%%%%%%%%%%%%%%%%%%%%%%%%%%%%%%%%%%%%%%%%%%%%%%%%%%%%%%%%%%%%%%%%%%%%%%%%%%%%%%%%%%%%

As discussed above, the faintest galaxies observed with GLIMPSE (\muv$\sim-14$) are a diverse sample, but few exhibit the extremely blue UV slopes and high escape fractions often assumed for faint galaxies in the EoR. To investigate whether this is consistent with simulations, we compare our results to the mock JWST observations of MEGATRON \citep{Katz2025}, a cosmological simulation with on-the-fly radiative transport that self-consistently models faint, high-z galaxies with sub-parsec resolution. 

To enable a direct comparison, we construct a mock sample of simulated galaxies broadly matched to the \muv\ distribution of our observed GLIMPSE sample. We bin GLIMPSE galaxies into 10 equal-width \muv\ intervals. From each \muv\ range, we randomly draw twice the number of simulated galaxies as observed with GLIMPSE, to sample a sufficient number of faint galaxies. The resulting mock sample spans a higher redshift range of $ 8 < z < 20 $, with a median redshift of $\langle z_{\rm M} \rangle \sim 10.3$ compared to GLIMPSE ($\langle z_{\rm G} \rangle \sim 6.7$). Both populations show minimal evolution in $\beta$ with redshift, justifying the comparison across redshift ranges.

%--Fig9: fesc-Muv simulations
\begin{figure*}
    \centering
    \includegraphics[width=\textwidth]{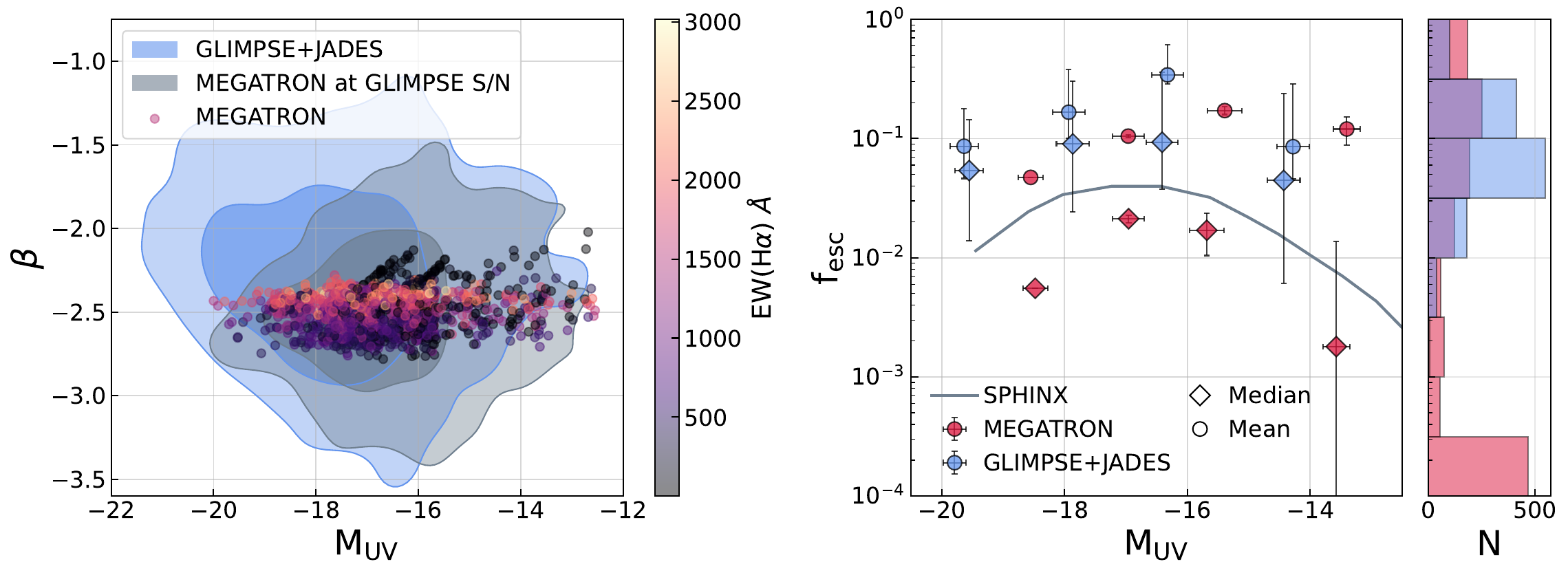}
    \caption{Comparison of faint galaxies from the MEGATRON simulation \citep{Katz2025} to GLIMPSE galaxies. \textbf{Left:} UV slope versus \muv\ for MEGATRON galaxies (points), colored by H$\alpha$ EW. Gray contours enclose the 68th and 95th percentiles of UV slopes measured from MEGATRON mock JWST photometry, perturbed to match GLIMPSE S/N at similar \muv. Blue contours enclose the GLIMPSE+JADES sample in Figure \ref{fig:beta-muv}. The observed spread in UV slopes is consistent with the MEGATRON points observed at GLIMPSE SNR. \textbf{Right:} LyC escape fraction for equal-width bins of \muv\ for MEGATRON (red points) and GLIMPSE (blue points). Both the median (diamonds) and mean (circles) \fesc\ are plotted, as the \fesc\ distribution is significantly skewed. The gray curve shows the z=6-7 mean \fesc\ from SPHINX simulations \citep{Rosdahl2022}.}
    \label{fig:fesc-muv-sim}
\end{figure*}

In the left panel of Figure \ref{fig:fesc-muv-sim}, we compare the MEGATRON UV slopes (scatter points) to those observed in GLIMPSE (blue contours). GLIMPSE galaxies are, on average, redder than simulated ($\langle \beta_{\rm G} \rangle \sim -2.28$, $\langle \beta_{\rm M} \rangle \sim -2.49$), reflecting increased dust attenuation. We observe a much broader range of $\beta$ values, largely due to photometric scatter. For each MEGATRON galaxy, we perturb the mock JWST photometry to match the average S/N observed with GLIMPSE at similar \muv\ and measure the UV slope as outlined in Section \ref{sec:uv_slopes}. This results in a similarly broad range of $\beta$ at faint \muv\ (gray contours), with MEGATRON being slightly bluer than the GLIMPSE observations at \muv $\sim -15$. The observed faint galaxy population is similarly as diverse as simulated.

Both observed and simulated galaxies display a wide range of H$\alpha$ EWs across all magnitudes (color-coding of points in left panel of Figure \ref{fig:fesc-muv-sim}), suggesting that faint, high-redshift galaxies can be actively star-forming or quiescent. The faint simulated galaxies show slightly redder UV slopes and a wider dispersion than their brighter counterparts. Notably, the majority of faint galaxies in MEGATRON exhibit low ionizing photon output ($\langle$\fesc$\times \xi_{\rm ion}\rangle$), but for two distinct reasons. 

The first subset of faint simulated galaxies have low escape fractions, strong H$\alpha$ emission, and relatively red UV slopes for such high redshift galaxies ($\beta \sim -2.4$). These are the orange points on the left panel of Figure \ref{fig:fesc-muv-sim}. Among the 75 galaxies in our mock sample with \muv\ $> -16$ and H$\alpha$ EW $\geq$ 1500Å, the average escape fraction is below 0.1\%, as emphasized by the histogram of the right panel. As discussed in Section \ref{sec:faint}, we observe a comparable population in GLIMPSE (faint galaxies with moderately red UV slopes and strong H$\alpha$ emission) suggestive of vigorously star-forming galaxies with UV slopes dominated by nebular continuum or reddened by dust attenuation. These galaxies lie in the upper-right corner of the $\beta$-EW(H$\alpha$) plot in the left panel of Figure \ref{fig:xiion}. The escape fractions predicted by MEGATRON for these galaxies support the interpretation that extreme nebular line-emission in faint, red galaxies requires low LyC escape fractions. 

In addition, we also identify faint MEGATRON galaxies with low ionizing photon output due to old stellar populations. The reddest UV slopes in our mock sample are from galaxies with evolved stellar populations which age toward fainter magnitudes and reduced H$\alpha$ emission, similar to the GLIMPSE galaxies shown in the middle column of Figure \ref{fig:faint} and those in the $z\sim6.3$ photometrically selected sample with upper limits on their H$\alpha$ emission (arrows in Figure \ref{fig:xiion}). Although these galaxies have high \fesc\ reported from MEGATRON, they no longer host the young, massive stars capable of producing ionizing photons and thus have a negligible contribution to reionization. 

Lastly, we do find evidence in MEGATRON for a subset of faint galaxies that contribute non-negligibly to reionization. The faint galaxies with the bluest UV slopes ($\beta < -2.6$) show weakened nebular emission ($\langle$EW H$\alpha \rangle$ = 427Å) due to their relatively high escape fractions ($\langle$\fesc$\rangle$=$28.9\%$). However, these blue and high \fesc\ galaxies comprise only $\sim$16\% of all galaxies fainter than \muv = $-16$ in the mock sample, indicating that high LyC leakage is not common among the faintest galaxies. We similarly observe few faint GLIMPSE galaxies with blue UV slopes and low H$\alpha$ emission needed for high \fesc. 

Overall, the majority of faint galaxies in the MEGATRON sample do not have high escape fractions, in agreement with our results for faint GLIMPSE galaxies. As shown by the histogram in the right panel of Figure \ref{fig:fesc-muv-sim}, the distribution of simulated \fesc\ is strongly skewed, with a long tail towards low values and a pileup at zero \fesc. As such, we plot both the MEGATRON median (red diamonds) and mean (red circles) escape fractions in equal-width bins of \muv. In the faintest bin, 53.8\% of simulated galaxies have \fesc\ $ < 0.1\%$, compared to only $\sim37\%$ in the brighter bins. As with GLIMPSE galaxies, both the median and mean escape fractions decrease towards the faintest bin. 

Additionally, we plot the mean escape fractions from the SPHINX simulation at z=6-7 \citep{Rosdahl2022} as the solid gray line in the right panel of Figure \ref{fig:fesc-muv-sim}. The predicted mean escape fraction peaks at \muv\ $\sim -17$, before sharply decreasing towards fainter galaxies. \citet{Rosdahl2022} attribute this peak to the mass range that is low enough for stellar feedback to remain effective, yet high enough for clustered star formation to produce concentrated supernovae which clear channels for LyC escape. 

In general, the escape fractions calculated in MEGATRON and SPHINX simulations agree with our estimated $\langle$\fesc$\rangle$ for the faint galaxies observed with GLIMPSE. This suggests that the faintest galaxies may not have the highest \fesc, providing crucial constraints for cosmic reionization.

\subsection{Cosmic Reionization with GLIMPSE}
%%%%%%%%%%%%%%%%%%%%%%%%%%%%%%%%%%%%%%%%%%%%%%%%%%%%%%%%%%%%%%%%%%%%%%%%%%%%%%%%%%%%%%%%%%%%%%%%%%%%%%%%%%%%%%%%%%%%%%%%%%%%%%%

Many studies suggest the faintest galaxies are the primary drivers of cosmic reionization due to their theoretically high number densities, large ionizing efficiencies, and high escape fractions as extrapolated from lower redshift studies of brighter galaxies. However, it is only recently with JWST that we are able to directly observe these faint galaxies in the EoR, and only with ultra-deep imaging of strongly lensed fields that we can observe statistical samples of the faintest galaxies. As such, we analyze what GLIMPSE observations of these faint, high-z galaxies imply about the progression and drivers of reionization. 

%--Fig10: nion vs. z 
\begin{figure}[h]
    \centering
    \includegraphics[width=\linewidth]{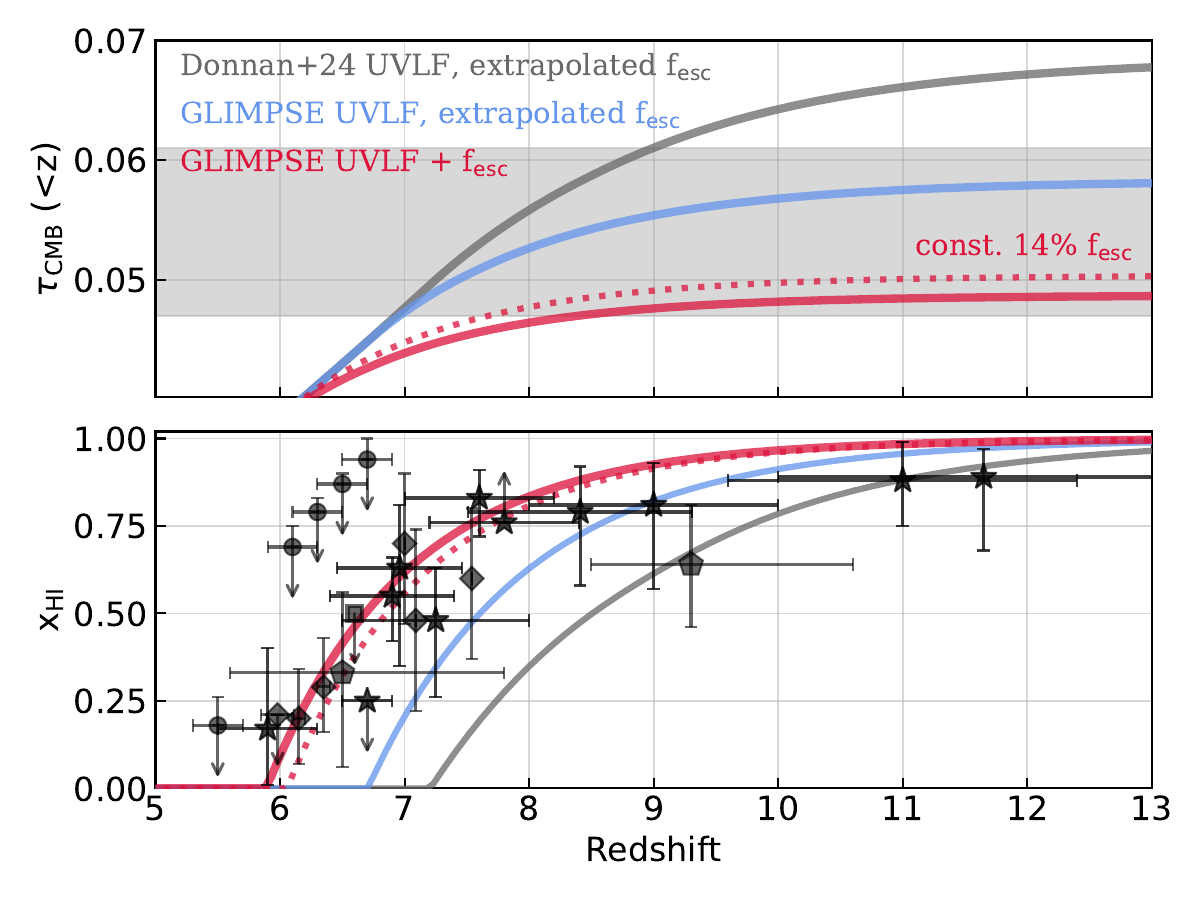}
    \caption{Cosmic reionization as observed with GLIMPSE (red lines) agrees with constraints on CMB optical depth \citep[gray shaded region, top;][]{Planck2016} and neutral hydrogen fraction (x$_{\rm HI}$) evolution with redshift \citep[bottom;][]{Mason2025}. The gray line uses the UVLF of \citet{Donnan2024} at z $>$ 9 and \citet{Bouwens2021} at z $\leq$ 9, while red and blue lines use that of \citet{Chemerynska2025}. The gray and blue lines use the \fesc\ prescription based on the UV slopes of \citet{Zhao2024} extrapolated to fainter galaxies (yellow lines in Figure \ref{fig:fesc-muv}), based on the low-z \fesc\ used in \citet{Munoz2024}. Solid and dotted red lines use GLIMPSE SDPL and constant 14\% \fesc(\muv), respectively. All models use the evolution of $\xi_{\rm{ion}}$ with \muv\ from \citet{Simmonds2024}. Code implementing this reionization model is available at \url{https://github.com/JulianBMunoz/Simple-Reionization-Plot}.}
    \label{fig:nion_z}
\end{figure}

We run the simple reionization model of \citet{Munoz2024} which solves for the volume-averaged neutral and ionized hydrogen fractions by integrating over the number density of ionizing photons from star-forming galaxies ($\dot n_{\rm ion} = \int dM_{\rm UV} \Phi_{\rm UV} \dot N_{\rm ion}f_{\rm esc}$). We use a clumping factor of C = 3 and integrate down to a cutoff magnitude of M$_{\rm UV}^{\rm ion. cutoff}$ = -12, the faintest GLIMPSE galaxy. The estimated ionizing photons of GLIMPSE galaxies and its evolution with time must balance recombinations in the IGM to match independent observational constraints on the CMB optical depth \citep{Planck2016} and neutral hydrogen fraction (x$_{\rm{HI}}(\rm z)$) \citep{Mason2025}. 

In Figure \ref{fig:nion_z}, we test how well various reionization histories compare to these observational constraints. The gray curve adopts the UVLF of \citet{Donnan2024} at z$>$9 and \citet{Bouwens2021} at z $\leq$ 9. The red and blue curves instead use the \citet{Chemerynska2025} double power-law UVLF fit to the z $>$ 8 GLIMPSE galaxies in our sample, evolving only the normalization with redshift as the faint-end slope and characteristic magnitude remain relatively constant. Both the gray and blue curves adopt the extrapolated \fesc(\muv,z) relation shown in Figure \ref{fig:fesc-muv}. The red curves show results using escape fraction prescriptions in agreement with GLIMPSE observations: the solid line uses the SDPL \fesc(\muv) fit, while the dotted line assumes a constant \fesc = 14\% for all \muv. All models use the z $\sim$ 6.5 $\xi_{\rm ion}$(\muv) from the mass-complete sample of \citet{Simmonds2024} which finds a modest redshift and \muv\ evolution with log($\xi_{\rm ion}) \simeq 25.4$. This relation agrees with the $\xi_{\rm ion}$ measurements calculated from the completeness-corrected z $\sim$ 6.3 GLIMPSE photometrically selected sample from Chisholm et al.\ (in prep.). The red curves therefore represent a self-consistent model in which ionizing parameters are directly informed by GLIMPSE observations of the faint galaxies in this sample. 

The difference between the blue and gray curves arises from the $\sim 0.5$ dex lower normalization of the \citet{Chemerynska2025} UVLF compared to that of \citet{Donnan2024}, as both have faint-end slopes of $\alpha \simeq -2.1$. Although the individual measurements of \citet{Donnan2024} are consistent with the \citet{Chemerynska2025} fit within uncertainties, the \citet{Donnan2024} extrapolated fit across redshift overpredicts the number of faint galaxies compared to GLIMPSE. Thus, the GLIMPSE UVLF predicts fewer faint galaxies at high redshift, resulting in a midpoint of reionization that happens at a $\Delta z \sim 0.5$ later in cosmic time.

The offset of the red curves from the blue is driven by the lower average \fesc\ inferred here for faint galaxies in GLIMPSE compared to previous extrapolations. The GLIMPSE SDPL fit predicts faint galaxies have a factor of 3 lower \fesc\ than previous predictions, slowing reionization by an additional $\Delta z \sim 0.75$ later in cosmic time. While the SPDL is the best-fit to robust GLIMPSE+JADES data, a constant \fesc = 14\% remains plausible given the uncertainties, and the prediction reionization history is shown by the dotted red line. We find that adopting the SDPL versus the constant \fesc\ only marginally changes the reionization history. 

The solid red line represents the reionization history informed by GLIMPSE observations of the faintest galaxies. This model agrees with the lower-end CMB optical depth constraint and nearly all Lyman-$\alpha$ forest x$_{\rm{HI}}$ constraints \citep{Mason2025}, unlike the blue and gray curves which use \fesc\ prescriptions extrapolated from brighter galaxies. The GLIMPSE model predicts reionization is halfway complete (x$_{\rm{HI}} = 0.5$) at z $\sim 6.7$ with a late ending of reionization \citep{Bosman2022, Becker2021}. The moderation of the reionization history largely arises from the GLIMPSE observations of the faintest galaxies that suggest modest \fesc\ and $\xi_{\rm{ion}}$, reducing \nion\ and slowing reionization to match x$_{\rm{HI}}(\rm z)$ observations. These results resolve the ionizing photon budget tension of \citet{Munoz2024} while still integrating down to feasible faint-end cutoff of $-12$. 

%--Fig11: nion vs. z
\begin{figure}[h]
    \centering
    \includegraphics[width=\linewidth]{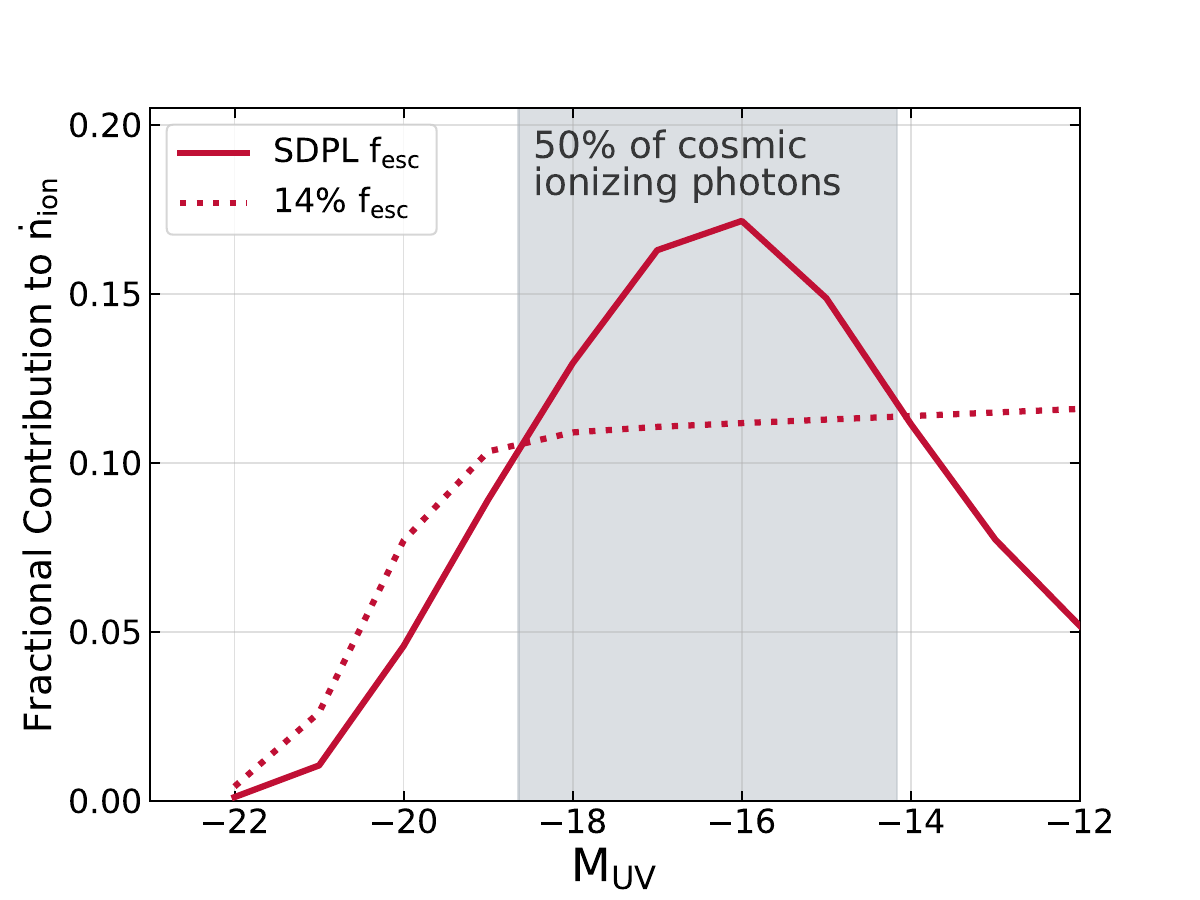}
    \caption{The contribution of ionizing photons to reionization as a function of \muv, normalized by the total ionizing photon budget over observed GLIMPSE galaxies ($\int^{-12}_{-22} \Phi_{\rm{UV}}$ $\xi_{\rm{ion}}$ \fesc\ L$_{\rm UV}$ d\muv). Both lines use GLIMPSE inputs of the corresponding models in Figure \ref{fig:nion_z}. The solid line uses the best-fit SDPL fit to \fesc(\muv) shown in Figure \ref{fig:fesc-muv}, predicting moderately faint galaxies with \muv $\sim -16$ are the drivers of reionization. The dashed line uses a constant 14\% \fesc, consistent with GLIMPSE findings.}
    \label{fig:nion_muv}
\end{figure}

Finally, in Figure \ref{fig:nion_muv} we determine what these updated $\Phi_{\rm UV}$, $\xi_{\rm ion}$, and \fesc\ prescriptions suggest about which galaxies were the drivers of reionization. We calculate the total ionizing photon contribution to reionization by integrating over the GLIMPSE $\Phi_{\rm UV}$ x \fesc\ x $\xi_{\rm ion}$ x L$_{\rm UV}$ and normalizing over our sample brightness range (\muv\ $\in$ [$-22,-12$]). The red solid and dashed lines use GLIMPSE-calibrated prescriptions from the corresponding reionization models of Figure \ref{fig:nion_z}. The GLIMPSE SDPL \fesc\ relation (solid line) peaks near \muv $\sim -16$ and drops off in both directions. While the steep UVLF does rise for fainter galaxies (faint-end slope $\alpha = -2.1$), it does not rise fast enough to compensate for \fesc\ and $\xi_{\rm{ion}}$ decreasing. This implies that galaxies with \muv\ between $-18$ and $-14$ mag were the drivers of cosmic reionization, a result consistent with the GLIMPSE \ion{O}{3} + H$\beta$ LF \citep{Kober25} and spectroscopic \fesc\ and \xiion\ predictions of \citet{Mascia2024} but slightly fainter than predicted in \citet{Matthee2022}. The SDPL relation places the first, second, and third quartiles of the cumulative \nion\ distribution at \muv $\simeq -18.2, -16.6,$ and $-15$, respectively, reinforcing that intermediate-luminosity systems produced the majority of ionizing photons. However, our large uncertainties on the \fesc\ of faint galaxies result in \fesc(\muv) also being consistent with a flat $\sim 14$\% \fesc. Shown as the dotted line in Figure \ref{fig:nion_muv}, this constant \fesc\ results in all galaxies in the faint-end of the UVLF (\muv $\geq -18$) contributing similar amount of ionizing photons to cosmic reionization.

GLIMPSE observations suggest galaxies around \muv $\sim -16$ are the prominent drivers of cosmic reionization, with galaxies fainter than \muv $\sim -14$ contributing only slightly or equally, at best. The surprising red UV slopes of faint GLIMPSE galaxies suggest a turnover in their \fesc\ below \muv $\sim -16$ resulting in a natural cutoff in the ionizing photon production where galaxies fainter than \muv $\sim -12$ do not significantly contribute to reionization.

%%%%%%%%%%%%%%%%%%%%%%%%%%%%%%%%%%%%%%%%%%%%%%%%%%%%%%%%%%%%%%%%%%%%%%%%%%%%%%%%%%%%%%%%%%%%%%%%%%%%%%%%%
\section{Conclusions}\label{sec:conclude}
%%%%%%%%%%%%%%%%%%%%%%%%%%%%%%%%%%%%%%%%%%%%%%%%%%%%%%%%%%%%%%%%%%%%%%%%%%%%%%%%%%%%%%%%%%%%%%%%%%%%%%%%%

Leveraging data from the ultra-deep, gravitationally lensed \textit{JWST}/NIRCam GLIMPSE survey \citep{Atek2025}, we present the first statistical sample of UV continuum slopes of the faintest galaxies in the EoR. This sample covers photometric redshifts of z = 6-16 and reaches absolute magnitudes down to \muv = $-12$, finding a total of 136 galaxies with \muv\ $> -16$. We estimate UV slopes by power-law fitting directly to filters in the rest-frame UV. We analyze UV slopes in combination with rest-optical emission from NIRCam medium-band filters to characterize faint galaxies, evaluate extremely blue UV slopes, and ultimately estimate LyC escape fractions to better understand cosmic reionization. We summarize our main findings below.

\begin{enumerate}%[itemsep=1ex, topsep=1ex]
  \item We find no statistically significant decrease in $\beta$ with redshift. We calculate d$\beta$/dz = $-0.031\pm0.016$, consistent with \citet{Topping2024, Robserts2024} but shallower than \citet{Cullen2024, Austin2024}. We find minimal difference in the redshift evolution of $\beta$ at fixed \muv. 
  
  \item The faintest galaxies (\muv$\gtrsim -16$) do not have the bluest UV slopes and are, at best, consistent with brighter galaxies. Our UV slopes of brighter (\muv $\leq -18$) galaxies agree with the best-fit trends of \citet{Cullen2023,Topping2024}, but extrapolating these trends towards fainter galaxies predicts much bluer slopes than observed with GLIMPSE. We find a large variance in observed $\beta$ values across all \muv. 
    
  \item We observe a diverse population of faint galaxies which includes dusty galaxies, older stellar populations, and low-mass star-forming galaxies. We find evidence for strong dust attenuation in low-mass galaxies as early as z $\sim$ 6.5. In general, galaxies with \muv $\gtrsim -16$ are not a monolithic population.
  
  \item We do not find convincing evidence for a population of galaxies with extremely blue UV slopes, even with the increased depth of GLIMPSE. Of the 52 galaxies with $\beta \leq -2.8$, most lie near the detection threshold and can be explained by photometric scatter. We classify only 6 galaxies as having more robust extreme $\beta$ values, but spectroscopy is needed to verify. 
  
  \item We analyze the relation between $\beta$, H$\alpha$ EW, \fesc, and A$_{\rm V}$ for the z$\sim$6.3 photometrically selected subsample where F480M covers H$\alpha$. We see that galaxies with red UV slopes typically have high dust attenuation and/or strong H$\alpha$ emission, indicating low LyC escape fractions. In comparison, galaxies with blue UV slopes require low dust attenuation and do not have the highest H$\alpha$ EW, consistent with higher escape fractions. We find similar behavior in the MEGATRON simulation, where faint galaxies have slightly redder UV slopes and low average ionizing output, justifying using the UV slope to estimate average \fesc\ for populations of faint galaxies at high redshifts. 
  
  \item The faintest galaxies do not have the highest escape fractions, as estimated from the $\beta$-\fesc\ relation of \citet{Chisholm2022}. We find that \fesc(\muv) is consistent with a constant mean $\sim 14\%$ \fesc, but best-fit by a smooth double power-law which peaks at \fesc$\sim$20\% at \muv=$-16.25$ and decreases for fainter galaxies. 

  \item We run the simple reionization model of \citet{Munoz2024} using the GLIMPSE UVLF of \citet{Chemerynska2025}, the $\xi_{\rm ion}$ distribution of \citet{Simmonds2024} which is consistent with GLIMPSE findings (Chisholm et al. in prep.), and \fesc\ from this work. These GLIMPSE observations of the ionizing properties of galaxies in the EoR produce a reionization history consistent with the CMB optical depth and Lyman-$\alpha$ forest constraints. Our best-fit \fesc\ results in galaxies between \muv $\simeq -18$ and $-14$ contributing $\sim 60\%$ of the ionizing photons need to reionization the universe. 
\end{enumerate}

Estimating LyC escape fractions for faint galaxies in the EoR is notoriously difficult. In general, GLIMPSE observations suggest that the faintest galaxies are a diverse population which, as a whole, do not necessarily have the highest \fesc. This initial GLIMPSE into the UV slopes of the faintest galaxies has provided a nuanced view of cosmic reionization and has provided new benchmarks for understanding the sources responsible.

\section*{acknowledgements}

We thank the referee for the insightful comments and suggestions that improved the clarity of this work. 
This work is based on observations made with the NASA/ESA/CSA James Webb Space Telescope. The data were obtained from the Mikulski Archive for Space Telescopes at the Space Telescope Science Institute, which is operated by the Association of Universities for Research in Astronomy, Inc., under NASA contract NAS 5-03127 for JWST. These observations are associated with programs \#3293 and \#9223. Support for programs \#3293 and \#9223 was provided by NASA through a grant from the Space Telescope Science Institute, which is operated by the Association of Universities for Research in Astronomy, Inc., under NASA contract NAS 5-03127. The data used are publicly available at the Mikulski Archive for Space Telescope (MAST), and can be accessed at \dataset[DOI: 10.17909/m2z4-t264]{https://doi.org/10.17909/m2z4-t264} and \dataset[doi:10.17909/8642-1k68]{https://doi.org/10.17909/8642-1k68}.

This work has received funding from the Swiss State Secretariat for Education, Research and Innovation (SERI) under contract number MB22.00072, as well as from the Swiss National Science Foundation (SNSF) through project grant 200020\_207349.  The Cosmic Dawn Center (DAWN) is funded by the Danish National Research Foundation under grant DNRF140. AA acknowledges support by the Swedish research council Vetenskapsr{\aa}det (VR) project 2021-05559, and VR consolidator grant 2024-02061. The Dunlap Institute is funded through an endowment established by the David Dunlap family and the University of Toronto. We acknowledge the support of the Canadian Space Agency (CSA) [25JWGO4A06].

\section{Appendix A}
Here we present a portion of the machine readable table (Table \ref{tab:catalogue}).

\begin{deluxetable*}{lcccccccccccccccc}
\rotate
\tablecaption{The GLIMPSE sample and derived properties used in this work.}
\label{tab:catalogue}

\tablehead{
\colhead{ID} &
\colhead{RA} &
\colhead{Dec.} &
\colhead{$\mu$} &
\colhead{M$_{\rm UV}$} &
\colhead{z$_{\rm phot}$} &
\colhead{$\beta_{\rm PL}$} &
\colhead{$\beta_{\rm PL,16th}$} &
\colhead{$\beta_{\rm PL,84th}$} &
\colhead{$\beta_{\rm SED}$} &
\colhead{$\beta_{\rm SED,16th}$} &
\colhead{$\beta_{\rm SED,84th}$} &
\colhead{$\beta_{\rm fesc}$} &
\colhead{$\beta_{\rm fesc,16th}$} &
\colhead{$\beta_{\rm fesc,84th}$} &
\colhead{A$_{\rm V}$}
}

\startdata
9948 & 342.194885 & -44.552319 & 1.78 & -16.30 $\pm$ 1.41 & 6.80 $\pm$ 0.27 & -2.07 & -2.40 & -1.74 & -2.04 & -2.24 & -2.08 & -2.29 & -2.34 & -2.10 & 0.23 \\
10400 & 342.182617 & -44.551617 & 2.18 & -16.21 $\pm$ 0.03 & 6.84 $\pm$ 0.64 & -1.16 & -2.17 & -0.96 & -1.95 & -2.15 & -2.04 & -2.18 & -2.27 & -1.97 & 0.23 \\
12380 & 342.178680 & -44.548706 & 2.84 & -15.43 $\pm$ 0.10 & 6.43 $\pm$ 0.23 & -2.46 & -2.91 & -2.01 & -2.32 & -2.30 & -2.06 & -2.31 & -2.48 & -2.10 & 0.00 \\
12786 & 342.181732 & -44.548149 & 2.78 & -15.56 $\pm$ 0.12 & 7.22 $\pm$ 0.26 & -2.28 & -3.25 & -2.06 & -2.16 & -2.24 & -2.09 & -2.34 & -2.49 & -2.10 & 0.00 \\
16257 & 342.181000 & -44.544476 & 4.35 & -15.50 $\pm$ 0.10 & 7.75 $\pm$ 0.32 & -1.80 & -2.04 & -1.56 & -2.02 & -2.14 & -1.97 & -1.99 & -2.13 & -1.97 & 0.00 \\
16271 & 342.183441 & -44.544506 & 3.95 & -15.22 $\pm$ 0.52 & 6.68 $\pm$ 0.36 & -1.43 & -2.05 & -0.87 & -2.10 & -2.17 & -2.07 & -1.89 & -2.32 & -1.95 & 0.00 \\
17876 & 342.175629 & -44.507442 & 1.96 & -16.54 $\pm$ 0.09 & 6.56 $\pm$ 0.21 & -2.11 & -2.44 & -1.78 & -2.01 & -2.14 & -1.97 & -2.18 & -2.23 & -1.99 & 0.00 \\
19355 & 342.170410 & -44.528893 & 12.41 & -13.89 $\pm$ 0.10 & 6.60 $\pm$ 0.26 & -1.98 & -2.44 & -1.50 & -2.12 & -2.21 & -2.09 & -2.33 & -2.37 & -2.08 & 0.00 \\
19881 & 342.205902 & -44.529102 & 4.95 & -15.02 $\pm$ 0.52 & 6.58 $\pm$ 0.19 & -1.91 & -2.30 & -1.52 & -2.18 & -2.27 & -2.10 & -2.41 & -2.42 & -2.09 & 0.06 \\
20025 & 342.209473 & -44.529198 & 3.38 & -16.37 $\pm$ 0.08 & 6.62 $\pm$ 0.11 & -2.27 & -2.46 & -2.10 & -1.91 & -2.08 & -1.93 & -2.02 & -2.20 & -2.01 & 0.00 \\
20026 & 342.209503 & -44.529259 & 3.34 & -16.45 $\pm$ 0.06 & 6.39 $\pm$ 0.08 & -2.20 & -2.34 & -2.07 & -1.92 & -2.00 & -1.87 & -2.07 & -2.13 & -2.01 & 0.00 \\
20891 & 342.190430 & -44.529491 & 3.53 & -16.35 $\pm$ 0.56 & 6.63 $\pm$ 0.54 & -3.34 & -3.78 & -1.75 & -2.38 & -2.31 & -2.09 & -2.66 & -2.49 & -2.08 & 0.00 \\
23217 & 342.157562 & -44.530296 & 3.03 & -15.32 $\pm$ 0.08 & 6.75 $\pm$ 0.29 & -2.13 & -2.49 & -1.71 & -2.12 & -2.17 & -2.06 & -1.91 & -2.32 & -1.99 & 0.00 \\
23448 & 342.197845 & -44.530403 & 34.16 & -12.53 $\pm$ 0.34 & 6.70 $\pm$ 0.17 & -2.12 & -2.41 & -1.79 & -2.16 & -2.14 & -2.05 & -2.15 & -2.30 & -2.02 & 0.00 \\
24483 & 342.193878 & -44.530766 & 13.14 & -13.98 $\pm$ 0.11 & 6.67 $\pm$ 0.25 & -2.23 & -2.72 & -1.56 & -2.15 & -2.17 & -2.07 & -2.11 & -2.32 & -2.01 & 0.01 \\
24893 & 342.158356 & -44.530888 & 3.40 & -15.61 $\pm$ 0.09 & 6.89 $\pm$ 0.19 & -1.91 & -2.27 & -1.62 & -2.00 & -2.17 & -2.06 & -2.14 & -2.33 & -2.05 & 0.00 \\
26321 & 342.189117 & -44.531368 & 12.86 & -14.16 $\pm$ 0.11 & 7.21 $\pm$ 0.43 & -1.97 & -2.30 & -1.12 & -2.08 & -2.20 & -2.08 & -1.93 & -2.37 & -1.98 & 0.03 \\
26864 & 342.190552 & -44.531559 & 12.77 & -14.21 $\pm$ 0.04 & 6.63 $\pm$ 0.77 & -1.51 & -2.04 & -1.20 & -2.01 & -2.16 & -2.06 & -1.89 & -2.26 & -1.95 & 0.13 \\
27601 & 342.189606 & -44.531796 & 15.16 & -14.27 $\pm$ 0.21 & 7.02 $\pm$ 0.33 & -1.82 & -2.48 & -1.49 & -2.12 & -2.21 & -2.08 & -2.30 & -2.44 & -2.14 & 0.00 \\
29612 & 342.189026 & -44.532528 & 24.74 & -14.52 $\pm$ 0.15 & 6.71 $\pm$ 1.14 & -2.08 & -3.05 & -0.86 & -2.11 & -2.16 & -2.04 & -1.94 & -2.29 & -1.96 & 0.00 \\
31042 & 342.171234 & -44.533058 & 6.55 & -14.56 $\pm$ 1.83 & 6.71 $\pm$ 0.76 & -1.71 & -2.81 & -1.72 & -1.84 & -2.23 & -2.07 & -1.98 & -2.36 & -2.04 & 0.29 \\
32505 & 342.206818 & -44.533699 & 2.85 & -16.70 $\pm$ 0.02 & 6.84 $\pm$ 0.02 & -2.49 & -2.65 & -2.34 & -2.15 & -2.17 & -2.11 & -2.40 & -2.42 & -2.31 & 0.01 \\
33643 & 342.208130 & -44.534206 & 2.67 & -15.88 $\pm$ 0.16 & 7.28 $\pm$ 0.27 & -3.07 & -3.44 & -2.39 & -2.35 & -2.36 & -2.12 & -2.67 & -2.59 & -2.18 & 0.00 \\
34689 & 342.164032 & -44.534698 & 19.03 & -13.93 $\pm$ 0.17 & 6.64 $\pm$ 0.15 & -1.83 & -2.19 & -1.46 & -2.06 & -2.16 & -2.06 & -2.26 & -2.35 & -2.07 & 0.02 \\
35742 & 342.189392 & -44.535172 & 20.15 & -13.91 $\pm$ 1.93 & 6.97 $\pm$ 0.43 & -1.78 & -2.15 & -1.55 & -1.95 & -2.21 & -2.09 & -2.03 & -2.31 & -2.06 & 0.31
\enddata

\tablecomments{Only the first 25 rows are shown. Full table available online, which also includes $M_{\rm b}$ and $M_{\rm c}$.}
\end{deluxetable*}

\bibliographystyle{aasjournal}
\bibliography{refs}

\end{document}